\documentclass[aps,prd,groupedaddress,twocolumn,nofootinbib]{revtex4-1}
\usepackage{graphicx}
\usepackage{amsmath}
\usepackage{amssymb}
\usepackage{bm}
\usepackage{float}
\usepackage[T1]{fontenc}
\usepackage{ae,aecompl}
\usepackage{color}
\usepackage{caption}
\usepackage{mathtools}
\usepackage{upgreek}

\pdfoutput=1

\newcommand{\bxi}{\boldsymbol{\xi}}

\begin{document}

\title{Tidal Effects in Eccentric Coalescing Neutron Star Binaries}

\author{Michelle Vick}
\email{mlv49@cornell.edu}
\affiliation{Cornell Center for Astrophysics and Planetary Science, Department of Astronomy, Cornell University, Ithaca, NY 14853, USA}
\author{Dong Lai}
\affiliation{Cornell Center for Astrophysics and Planetary Science, Department of Astronomy, Cornell University, Ithaca, NY 14853, USA}

\date{\today}

\begin{abstract}
Dynamically formed compact object binaries may still be eccentric while in the LIGO/Virgo band. For a neutron star (NS) in an eccentric binary, the fundamental modes (f-modes) are excited at pericenter, transferring energy from the orbit to oscillations in the NS. We model this system by coupling the evolution of the NS f-modes to the orbital evolution of the binary as it circularizes and moves toward coalescence. NS f-mode excitation generally speeds up the orbital decay and advances the phase of the gravitational wave signal from the system. We calculate how this effect changes the timing of pericenter passages and examine how the cumulative phase shift before merger depends on the initial eccentricity of the system. This phase shift can be much larger for
highly eccentric mergers than for circular mergers, and can be used to probe the NS equation of state. 
\end{abstract}

\maketitle

\section{\label{sec:Intro} Introduction}

In its first and second observing runs, the LIGO/Virgo collaboration has detected 10 black hole (BH) binary mergers \cite{Ligo18} and one neutron star (NS) binary merger \cite{Ligo17}. As LIGO and Virgo improve in sensitivity, they are expected to detect many more NS binary merger events. The proposed formation channels for compact object (CO) binary mergers can be divided into two broad classes: isolated binary evolution and dynamical formation. In the first, an isolated stellar binary becomes tighter in separation due to drag forces in the common-envelope phase \cite[e.g.][]{Lipunov97,Lipunov17,Podsiadlowski03,Belczynski10,Dominik12,Dominik13,Dominik15,Belczynski16}. CO binaries that form via this pathway are expected to be circular while emitting gravitational waves (GWs) in the LIGO band. In the second class, CO binaries form dynamically through gravitational interactions between multiple stars and COs. For instance, BH binaries in dense star clusters can become bound and shrink in separation due to three-body encounters (e.g. an exchange interaction between a binary and a CO) and/or secular interactions \cite[e.g.][]{PZ00,Miller02,Wen03,OLeary09, Miller09,Antonini12,Rodriguez15,Samsing18}. Another type of dynamical formation occurs in the galactic field, where CO mergers are induced in hierarchical triple or quadruple systems \cite{Silsbee17,Antonini17,Liu18,Liu19,Liu19b}. Intriguingly, some fraction of dynamically assembled CO binaries may emit GWs within the LIGO band while their orbits are still highly eccentric. The formation rate for such binaries is uncertain, but detecting these eccentric systems by LIGO/Virgo would be of great interest.

The effects of tides on the gravitational waveform of coalescing NS binaries in circular orbits have been studied in many papers \cite[e.g.][]{Kochanek92,Bildsten92,Lai94a,Lai96b,Baumgarte98,Binnington09,Damour09,Uryu09,Penner11,Ferrari12}; (see Section I of \cite{Xu18} for a short review). 
An analytical expression for the GW phase shift due to quasi-equilibrium tides (f-mode distortion) was derived in \cite{Lai94} and \cite{Flanagan08}. The effects of resonant tides have also been explored \cite{Lai94b,Shibata94,Reisenegger94,Ho99,Lai06,Yu17a,Yu17b,Andersson18,Xu18,Yang19}. 

In this work, we study the effect of dynamical tides on the orbital decay and the resulting GW signal from an eccentric CO binary with a NS. We focus on the f-mode oscillation of the NS as other modes (g-modes and r-modes) couple rather weakly with the tidal potential and produce very small effects even in resonance with circular orbits (see \cite{Xu18} and references therein). By coupling the f-mode evolution to the post-Newtonian (PN) orbital evolution, we calculate the effect of tides on the GW signal as the binary evolves toward merger. Our model contributes to a growing body of analytical and numerical work on eccentric NS binaries \cite{Chirenti17,Parisi18, Yang18, Chaurasia18, Yang19} by accurately computing the amplitude of the NS f-mode as the binary decays and circularizes due to gravitational radiation. We study binaries with a range of initial pericenter separations and eccentricities to quantify how such dynamical tides affect orbital evolution as a function of eccentricity. 

In Section~\ref{sec:Theory}, we present our model for evolving the NS f-mode and binary orbit. In Section~\ref{sec:noGR} we discuss the behavior of the mode-orbit coupling in the absence of relativistic effects. In Section~\ref{sec:Results} we present results of our calculations for binaries with a range of initial pericenter separations and eccentricities, and we conclude in Section~\ref{sec:Discussion}. The Appendix contains an analytical assessment of the mode-orbit resonance effect, which we show generally produces a small GW phase shift.

\section{Equations of Motion Including Dynamical Tides and GR Effects \label{sec:Theory}}

		The orbit of a NS binary evolves in response to the tides raised on the NS as well as general relativity (GR). For a binary with a NS (mass $M_1$ and radius $R_1$) and companion $M_2$ (either another NS or a BH), the Newtonian gravitational potential produced on $M_1$ by $M_2$ is 
		\begin{equation}
		U(\boldsymbol{r},t) = - M_2\sum_{lm} \frac{W_{lm} r^l}{D(t)^{(l+1)}} \text{e}^{- \text{i} m \Phi(t)} Y_{lm} (\phi,\theta), \label{eq:potential}
		\end{equation}
		where $\textbf{r} = (r,\theta,\phi)$ is the position vector in spherical coordinates with respect to the center of mass of $M_1$, $D(t)$ and $\Phi(t)$ are respectively the orbital separation and true anomaly, and
		\begin{align}
		W_{lm} =& (-1)^{(l+m)/2}\left[\frac{4\uppi}{2l+1}(l+m)!(l-m)!\right]^{1/2} \nonumber \\ & \times \left[2^l\left(\frac{l+m}{2}\right)!\left(\frac{l-m}{2}\right)!\right]^{-1}.
		\end{align}
		We adopt units such that $G=c=1$ throughout the paper. We will focus on the dominant quadrupole tides ($l=2$), for which $W_{2\pm 2} = \sqrt{3\pi/10}$, $W_{2\pm 1} = 0$, and $W_{20} = \sqrt{\pi/5}$.
		
		The Lagrangian displacement vector $\boldsymbol{\xi}(\boldsymbol{r},t)$ denotes the fluid perturbation on $M_1$ driven by the tidal potential. We can decompose $\boldsymbol{\xi}(\boldsymbol{r},t)$ into normal modes $ \boldsymbol{\xi}_\alpha(\boldsymbol{r})\propto \text{e}^{\text{i} m \phi} $ of frequencies $\omega_\alpha$, where $\alpha = \{n_rlm\}$ specifies the mode index: 
		\begin{equation}
		\left[\begin{array}{c}
		\bxi\\
		{\partial\bxi/\partial t}
		\end{array}\right]
		=\sum_\alpha c_\alpha(t)
		\left[\begin{array}{c}
		\bxi_\alpha(\boldsymbol{r})\\
		-i\omega_\alpha\bxi_\alpha(\boldsymbol{r})
		\end{array}\right].
		\end{equation}
		A freely oscillating mode has $\boldsymbol{\xi}(\boldsymbol{r},t) \propto \text{e}^{\text{i} m \phi - \text{i} \omega_\alpha t}$. This decomposition includes both positive and negative mode frequencies \cite{Schenk02}. We neglect NS rotation and adopt the convention $\omega_\alpha>0$ such that $m>0$ corresponds to prograde modes and $m<0$ to retrograde modes. The mode amplitude $c_\alpha(t)$ satisfies 
		\begin{equation}
		\dot{c}_\alpha + i \omega_\alpha c_\alpha = \frac{i M_2 W_{lm}Q_\alpha}{2 \omega_\alpha D^{l+1}}\text{e}^{-\text{i}m \Phi},
		\label{eq:cdot}
		\end{equation}
		with
		\begin{equation}
		Q_\alpha \equiv \int d^3x \; \rho \boldsymbol{\xi}_\alpha^*\cdot\nabla(r^lY_{lm}) .\label{eq:defQ}
		\end{equation}
		In Eqs.~(\ref{eq:cdot}) and (\ref{eq:defQ}), $\bxi_\alpha$ is normalized such that $\langle \bxi_\alpha, \bxi_\alpha \rangle \equiv \int d^3x \; \rho \boldsymbol{\xi}_\alpha^*\cdot \boldsymbol{\xi}_\alpha = 1$, and we have adopted the units $G=M_1 = R_1$ in these equations, so that $Q_\alpha$ is dimensionless \cite{Lai06,Fuller12a}.
		
		The general relativistic equations of motion of compact binaries in eccentric orbits are rather complicated, and even the notion of eccentricity is difficult to define in general relativity \cite[e.g.][]{Blanchet14,Loutrel19}. For the purpose of our study, we find it convenient to use the effective one-body PN equations of motion developed by \cite{Lincoln90} (see also \cite{Kidder93} for discussion). These equations of motion contain all PN corrections through (Post)$^{5/2}$-Newtonian order, including effects due to the radiation reaction. We incorporate the tidal effect in the same way as in \cite{Fuller11}. Thus, restricting to the $l=2$ modes, the orbital evolution equations are
		\begin{align}
		\ddot{D} =& D\dot{\Phi}^2-\sum_{\alpha}\frac{3M_t}{D^4}W_{2m}Q_{\alpha} \left(\text{e}^{\text{i}m\Phi}c_{\alpha}\nonumber +\text{c.c.}\right) \\
		&- \frac{ M_t}{D^2}\left(1+ A_{\rm PN}+A_{5/2}+B_{\rm PN}\dot{D} + B_{5/2}\dot{D}\right) \label{eq:Dddot}, \\
		\ddot{\Phi} =& -\frac{2\dot{D}\dot{\Phi}}{D} + \sum_{\alpha}im\frac{M_t}{D^5} W_{2m}Q_{\alpha} \left(\text{e}^{\text{i}m\Phi}c_{\alpha} -\text{c.c.}\right)\nonumber \\ &-\frac{M_t}{D^2}\left(B_{\rm PN}+B_{5/2}\right)\dot{\Phi} \label{eq:Phiddot},
		\end{align}
		where the sum over $\alpha$ is restricted to positive mode frequencies (with $m=\pm2,0$) and $M_t = M_1+M_2$ is the total mass. Throughout this paper, we use the values $\omega_\alpha = 1.22 \; (M_1/R_1^3)^{1/2}$ and $Q_\alpha = 0.56$, which correspond the $l=2$ f-mode of a $\Gamma=2$ polytrope. In Eqs.~(\ref{eq:Dddot}) and (\ref{eq:Phiddot}), $A_{5/2}$ and $B_{5/2}$ represent the leading-order gravitational radiation reaction forces, and the $A_{\rm PN}$ and $B_{\rm PN}$ terms are the non-dissipative first and second-order PN corrections. These coefficients are given by
		\begin{align}
		A_{5/2} =& -\frac{8\mu}{5D}\dot{D}\left(18v^2 +\frac{2M_t}{3D} - 25\dot{D}^2\right),\\
		B_{5/2} =& \frac{8\mu}{5D}\left(6v^2 -\frac{2M_t}{D} - 15\dot{D}^2\right),\\
		A_{\rm PN} =& (1+3\eta)v^2-2(2+\eta)\frac{M_t}{D}-\frac{3}{2}\eta\dot{D}^2\nonumber \\ & + \frac{3}{4}(12+29\eta)\left(\frac{M_t}{D}\right)^2 + \eta(3-4\eta)v^4 \nonumber \\
		& + \frac{15}{8}\eta(1-3\eta)\dot{D}^4 - \frac{3}{2}\eta(3-4\eta)v^2\dot{D}^2 \nonumber \\
		& - \frac{1}{2}\eta(13-4\eta)\frac{M_t}{D}v^2 - (2+25\eta+2\eta^2)\frac{M_t}{D}\dot{D}^2,\\
		B_{\rm PN} =& -2(2-\eta)\dot{D} - \frac{1}{2}\dot{D}\left[\vphantom{\frac{M_t}{D}}\eta(15+4\eta)v^2 \right. \nonumber \\ & \left. -(4+41\eta+8\eta^2)\frac{M_t}{D} - 3\eta(3+2\eta)\dot{D}^2\right],
		\end{align}
		with $\mu = M_1M_2/M_t$ the reduced mass, $\eta = \mu/M_t$, and $v^2 = \dot{D}^2 +(D\dot{\Phi})^2$. 
		
		Note that while Eqs.~(\ref{eq:Dddot}) and (\ref{eq:Phiddot}) include gravitational radiation associated with the orbital motion, they do not include gravitational radiation due to the tidally excited oscillation modes. Incorporating the latter effect is complicated by the fact that the orbit and modes can radiate coherently (see \cite{Lai94b} for the circular orbit case where this effect can be included in an approximate way), and is beyond the scope of this paper. Because of this, our results in Section~\ref{sec:Results} underestimate the influence of dynamical tides on the orbital evolution.
		
		The total energy of the system is the sum of the energy in the oscillation modes and the orbital energy, including the interaction between the modes and the gravitational potential. The total energy in stellar oscillations is
		\begin{equation}
		E_{\rm mode} =2 \sum_{\alpha} \omega_{\alpha}^2 |c_{\alpha}|^2, \label{eq:defEmodes}
		\end{equation}
		where the sum is again restricted to positive mode frequencies. The Newtonian expression for the orbital energy is 
		\begin{align}
		E_{\rm orb} =& - \frac{\mu M_t}{D} + \frac{\mu}{2}\left(\dot{D}^2 + D^2\dot{\Phi}^2\right) \nonumber \\& -  \mu M_t \sum_{\alpha} \frac{W_{2m}Q_{\alpha}}{D^3}(\text{e}^{\text{i}m\Phi}c_{\alpha} +\text{c.c.}).\label{eq:defEorb}
		\end{align}
		When GR effects are neglected (i.e. $A_{\rm PN} = B_{\rm PN}=A_{5/2}=B_{5/2}=0)$, the total energy $E_{\rm tot} = E_{\rm orb} + E_{\rm mode}$ is conserved. 
\section{Orbit and Mode Evolution without GR\label{sec:noGR}}
\begin{figure}
	\includegraphics[width = 3in]{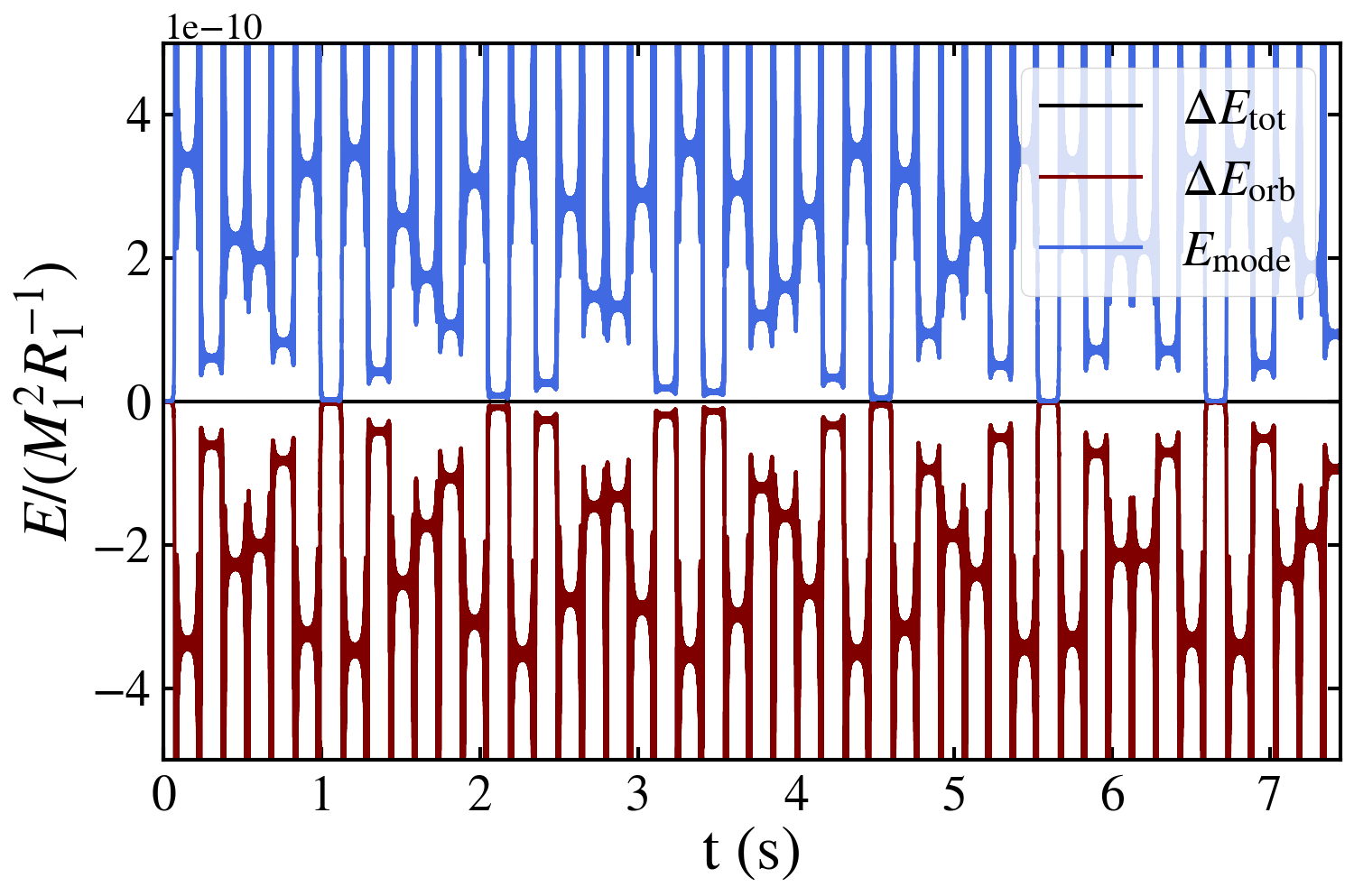}
	\caption{The evolution of the mode energy, Eq.~(\ref{eq:defEmodes}), orbital energy ($\Delta E_{\rm orb} = E_{\rm orb} - E_{\rm orb,0}$), Eq.~(\ref{eq:defEorb}), and total energy ($\Delta E_{\rm tot} = E_{\rm tot} - E_{\rm tot,0} = E_{\rm mode}+\Delta E_{\rm orb}$) in units where $G=M_1=R_1=1$ for a binary with a single NS. We have used a $\Gamma=2$ polytrope to model the NS with $\omega_\alpha = 1.22 \; (M_1/R_1^3)^{1/2}$ and $Q_\alpha = 0.56$. The initial pericenter and eccentricity are $D_{\rm p,0} = 5.995R_1$ and $e_0 = 0.9$, corresponding to $|\Delta \hat{P}_\alpha|=1.6\times 10^{-4}$ (see Eq.~\ref{eq:defDelPhat}). This calculation does not include GR (i.e. $A_{5/2} = B_{5/2} = A_{\rm PN} = B_{\rm PN} = 0$). The $l=2$, $m=(2,0,-2)$ f-modes are all accounted for in the integration. The mode energy undergoes small-amplitude oscillations over multiple orbits.}
	\label{fig:EnergyConsrp5p995}
\end{figure}
\begin{figure}
	\includegraphics[width = 3in]{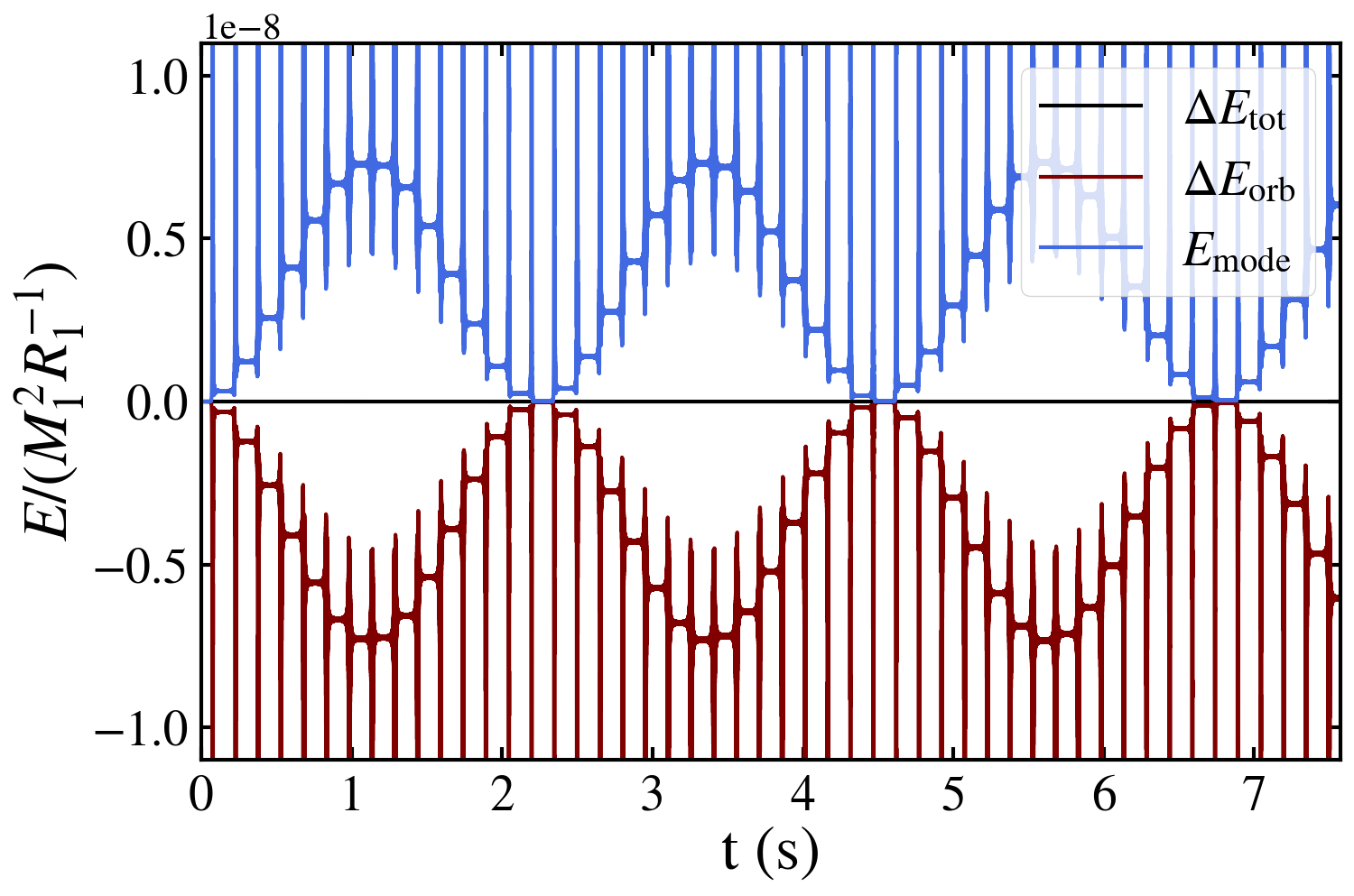}
	\caption{Same as Fig.~\ref{fig:EnergyConsrp5p995}, but with an initial pericenter distance of $D_{\rm p,0} = 6R_1$ and $|\Delta \hat{P}_\alpha|=1.5\times 10^{-4}$. The mode energy can reach larger values than in Fig.~\ref{fig:EnergyConsrp5p995} due to a resonance between the mode frequency and the orbital frequency  ($\omega_\alpha \simeq 401 \Omega_{\rm orb}$).} \label{fig:EnergyConsrp6}
\end{figure}
\begin{figure}
	\includegraphics[width = 3in]{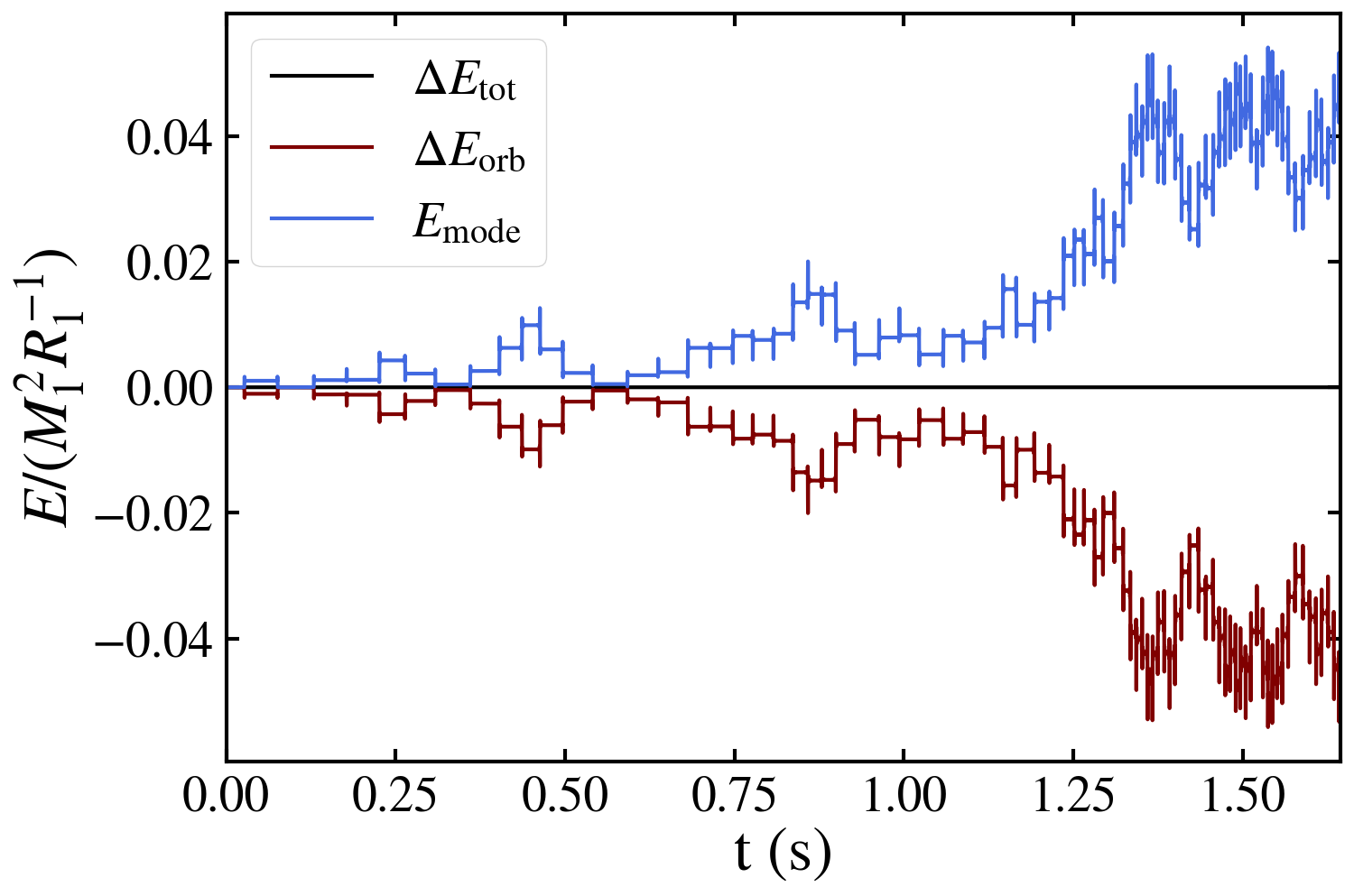}
	\caption{Same as Fig.~\ref{fig:EnergyConsrp5p995}, but with a smaller initial pericenter distance of $D_{\rm p,0} = 3R_1$ so that $|\Delta \hat{P}_\alpha|= 82$. The mode energy grows chaotically over many orbits. } \label{fig:EnergyConsrp3}
\end{figure}
	
	Before studying coalescing binaries using the full equations from Section \ref{sec:Theory}, we consider in this section the ``mode + orbit" problem without GR (i.e. we set $A_{\rm PN}=B_{\rm PN}=A_{5/2}=B_{5/2}=0$). Previous studies of dynamical tides in a variety of astrophysical situations \cite{Mardling95a, Lai96a, IP04,Vick18,Wu18,Vick19} have demonstrated that the coupled evolution of the eccentric orbit and tidally driven oscillation modes can yield different behaviors depending on the binary orbital properties. We briefly discuss how the binary pericenter distance $D_{\rm p}$ and eccentricity $e$ affect the interaction between the orbit and the oscillation modes. For a more thorough analysis, see Section 2 of \cite{Vick18}. 
		
		For a NS in an eccentric binary, the $l=2$ f-mode is excited most strongly at pericenter, and the mode amplitude changes by the real quantity $\Delta c_\alpha$ (see Eq. 10 of \cite{Vick18}) during each pericenter passage, transferring energy and angular momentum between the orbit and the  NS f-mode. When the binary is highly eccentric, the shape of the NS orbit near pericenter is unchanged over many orbits, and $\Delta c_\alpha$ remains constant over multiple pericenter passages. We can relate $\Delta c_\alpha$ to a change in the mode energy in the ``first" passage (i.e. when there is no pre-existing mode oscillation)
		\begin{equation}
		\Delta E_{\rm mode} = 2 \sum_{\alpha}\omega_\alpha^2(\Delta c_\alpha)^2.
		\end{equation}
		As the mode energy changes, so too will the orbital energy, causing a slight adjustment, $|\Delta P|$, in the initial orbital period (initially $P_0$). We define 
		\begin{equation}
		|\Delta \hat{P}_\alpha| \equiv \omega_\alpha|\Delta P| \simeq \frac{3}{2} \omega_\alpha P_0 \left(\frac{\Delta E_{\rm mode}}{|E_{\rm orb,0}|}\right), \label{eq:defDelPhat}
		\end{equation}
		where $E_{\rm orb,0}$ is the initial orbital energy. Physically, $|\Delta \hat{P}_\alpha|$ is the phase shift in the mode oscillation due to tidal energy transfer at pericenter. The phase shift is largest for binaries with strong tidal interactions (small $D_{\rm p}$) and large orbital periods (high $e$). 
		
		In the absence of mode damping and GR, the properties $|\Delta \hat{P}_\alpha|$ and $\omega_\alpha P_0$ determine the behavior of the ``mode + eccentric orbit" system over multiple orbits. The system exhibits three types of behavior:
		\begin{enumerate}
			\item When $|\Delta \hat{P}_\alpha| \lesssim 1$, the orbit and the f-mode oscillations gently trade a small amount of energy (of order $\Delta E_{\rm mode}$) back and forth, as shown in Fig.~\ref{fig:EnergyConsrp5p995}.
			\item When $|\Delta \hat{P}_\alpha| \lesssim 1$ and $\omega_\alpha P_0 = 2 \pi n$ (with integer $n$), the mode exhibits resonant behavior, with the mode energy climbing to $E_{\rm mode} \gg \Delta E_{\rm mode}$, but still undergoing oscillations (see Fig.~\ref{fig:EnergyConsrp6}).
			\item  When $|\Delta \hat{P}_\alpha| \gtrsim 1$, the mode energy grows chaotically and can reach an appreciable fraction of the NS binding energy (see Fig.~\ref{fig:EnergyConsrp3}). This behavior occurs because the pericenter energy transfer changes the orbital period enough that the phase of the f-mode at pericenter is nearly random from one orbit to the next. The chaotic mode growth resembles a diffusive process, except there exists an ``upper floor" that the mode energy can attain. Note that the linear mode treatment is no longer appropriate when the f-mode energy becomes too large.   
		\end{enumerate}

		A highly eccentric NS binary may pass through the regimes for all three behaviors --- low-amplitude oscillations, resonance, and chaotic growth --- as gravitational radiation shrinks the orbit. However, as we shall see in Section \ref{sec:Results} (see also the Appendix), because of the rapid orbital decay, these behaviors may not manifest as prominently as in the case of non-dissipative systems.
		
\section{Orbit and Mode Evolution Including GR}\label{sec:Results}

\begin{figure*}
	\begin{center}
		\includegraphics[width=6.5in]{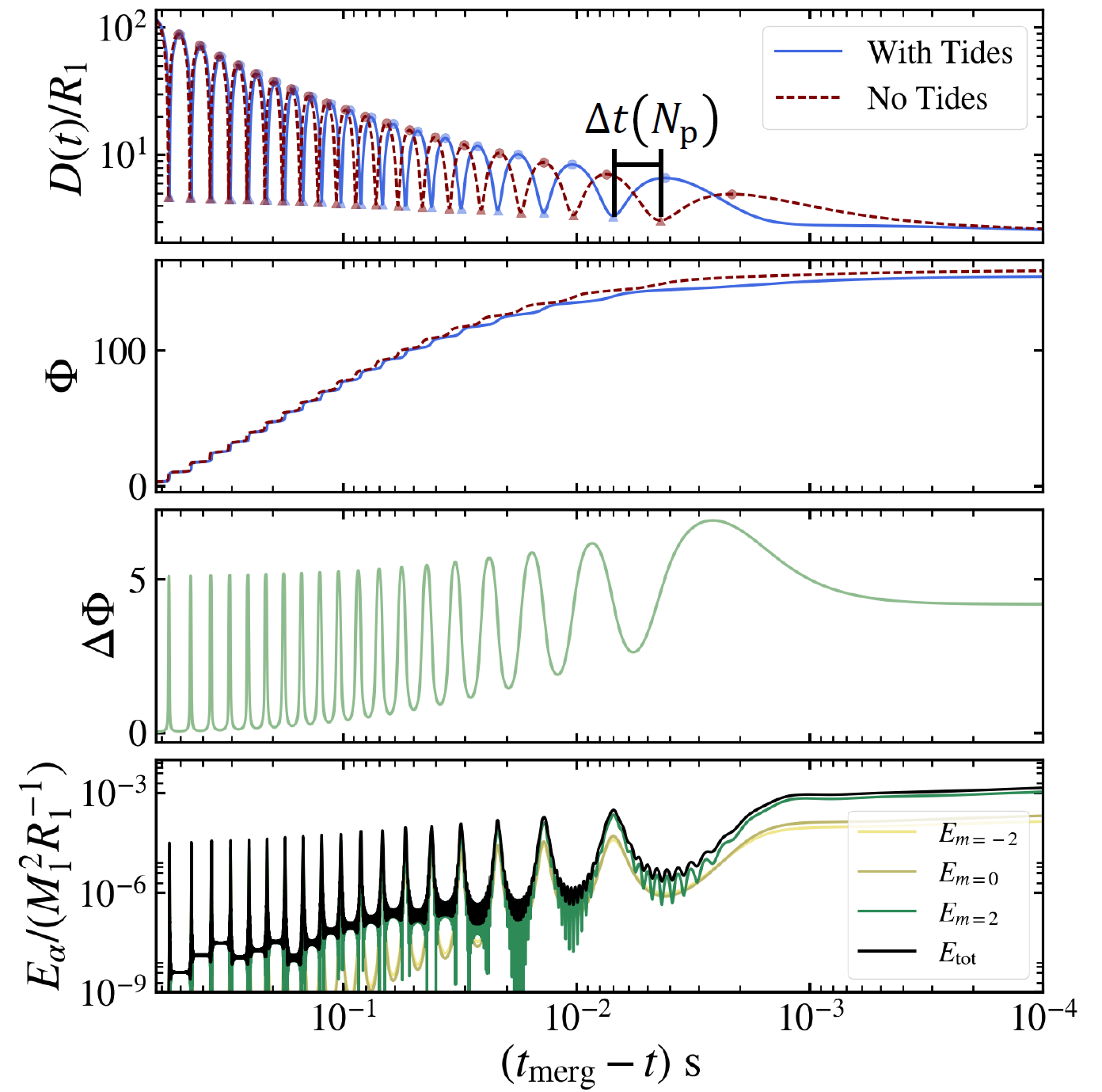}
		\caption{An example of how f-mode oscillations alter the orbital evolution of a coalescing, eccentric NS binary with $M_1=M_2=1.4M_\odot$ and $R_1=R_2=10$~km. The NSs are modeled as $\Gamma=2$ polytropes. This system has initial eccentricity $e_0=0.9$ and pericenter distance $D_{\rm p,0} = 6 R_1$, corresponding to a GW pericenter frequency of $f_{\rm p,0} = 575$~Hz. The solid blue lines in the top two panels show the binary separation and orbital phase (true anomaly) including tidal effects, while the dashed red lines show the same without tides. The blue (red) circles and triangles mark the times of apocenter and pericenter. The quantity $\Delta t(N_p)$ (shown in the top panel) is defined as the difference in the timing of a pericenter passage for calculations with and without tides. The third panel shows $\Delta \Phi$, the difference in the orbital phase for calculations with and without dynamical tides (see Eq.~\ref{eq:DeltaPhi}). The bottom panel shows the evolution of the mode energies; the $m=2$ (prograde) mode dominates.} \label{fig:OrbitalEvolution}
	\end{center}
\end{figure*}

\begin{figure*}
	\begin{center}
		\includegraphics[width=6in]{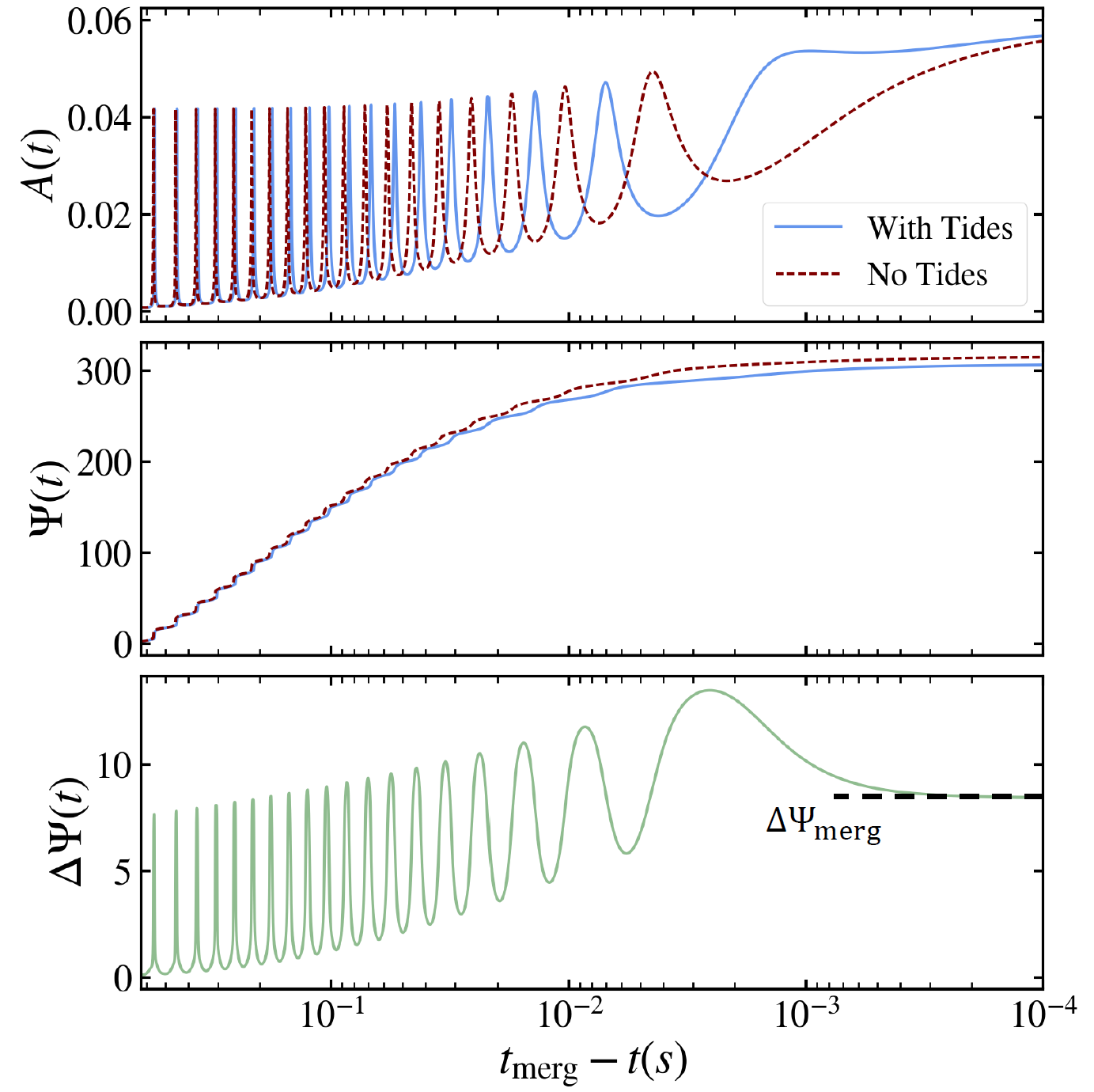}
		\caption{The gravitational waveform (Eq.~\ref{eq:APhidef}) that corresponds to the orbital evolution shown in Fig.~\ref{fig:OrbitalEvolution}. The amplitude is scaled by $(R_1/d)$, with $d$ the distance to the system. The bottom panel shows the difference in the phase of the GWs, $\Delta \Psi (t)$, for a calculation with dynamical tides and one without (Eq.~\ref{eq:DeltaPsi}).}
		\label{fig:waveform}
	\end{center}
\end{figure*}

\begin{figure*}
	\includegraphics[width=\columnwidth]{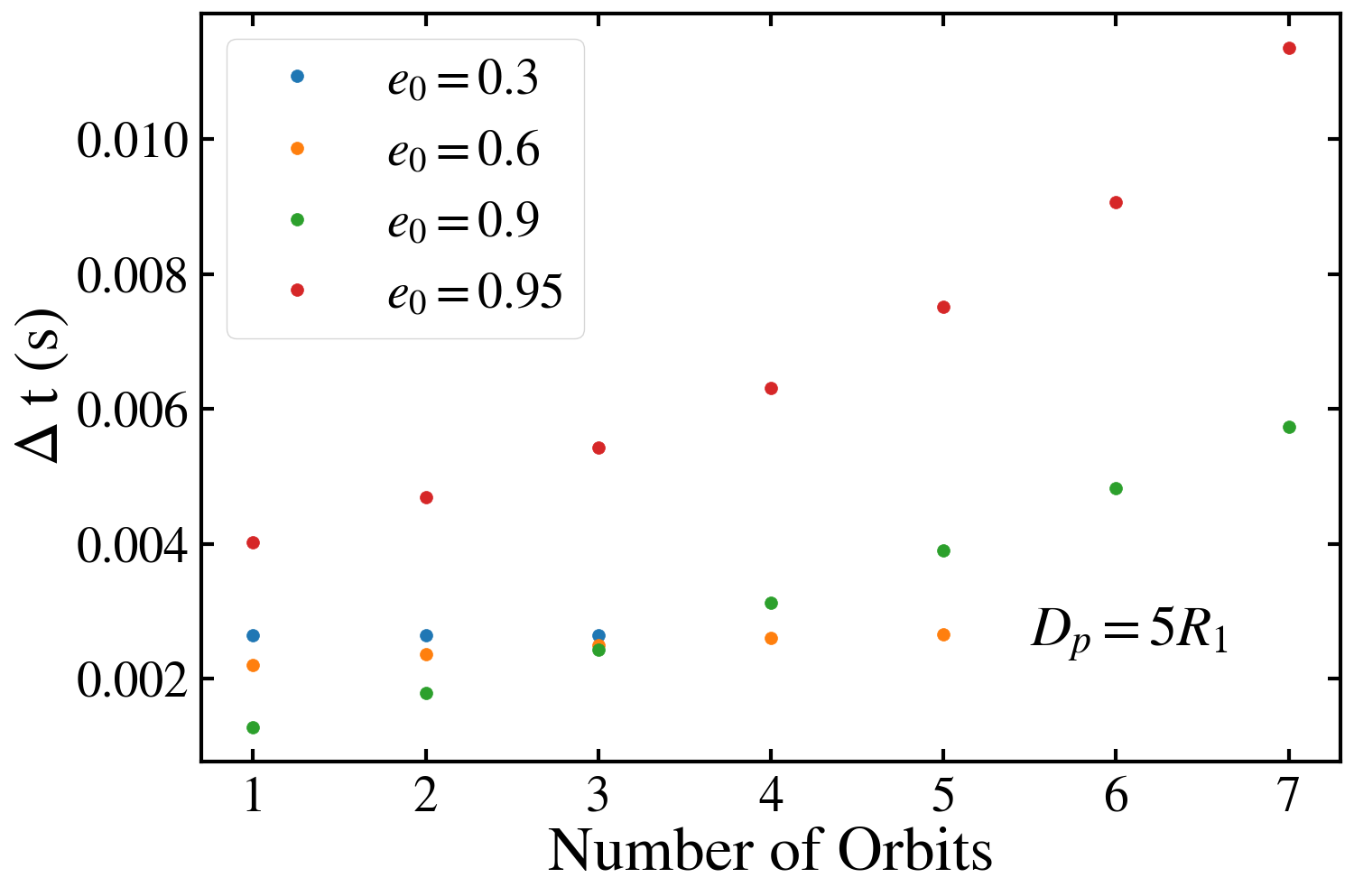}
	\includegraphics[width=\columnwidth]{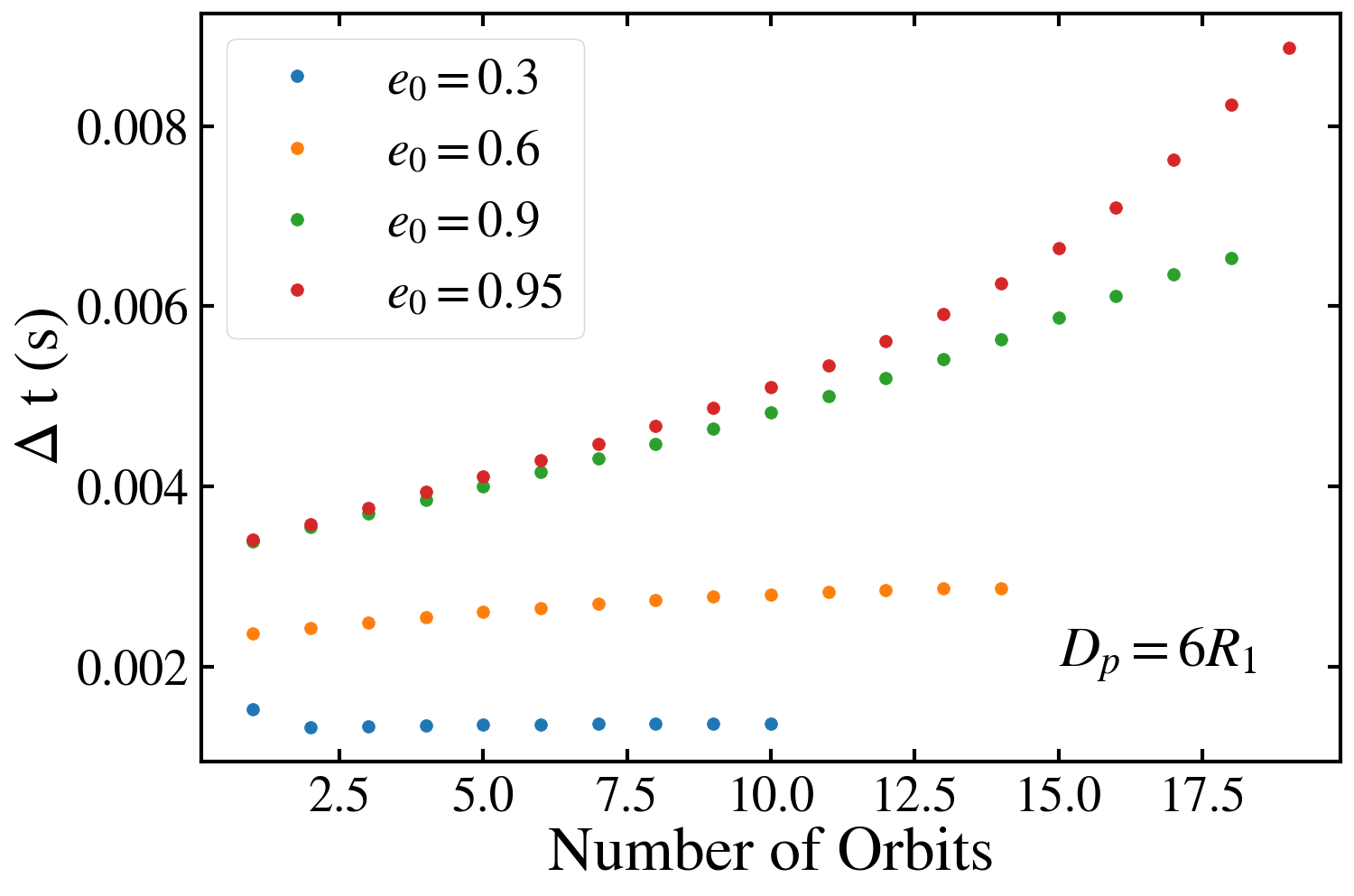}
	\caption{Cumulative time advance due to dynamical tides [see the top panel of Fig.~\ref{fig:OrbitalEvolution} for the definition of $\Delta t(N_p)$] as a function of the number of orbits for equal mass NS binaries with an initial pericenter distance of $D_{\rm p,0} = 5 R_1$ in the left panel and $D_{\rm p,0} = 6 R_1$ in the right panel.}
	\label{fig:Deltat}
\end{figure*}

\begin{figure*}
	\includegraphics[width = 6in]{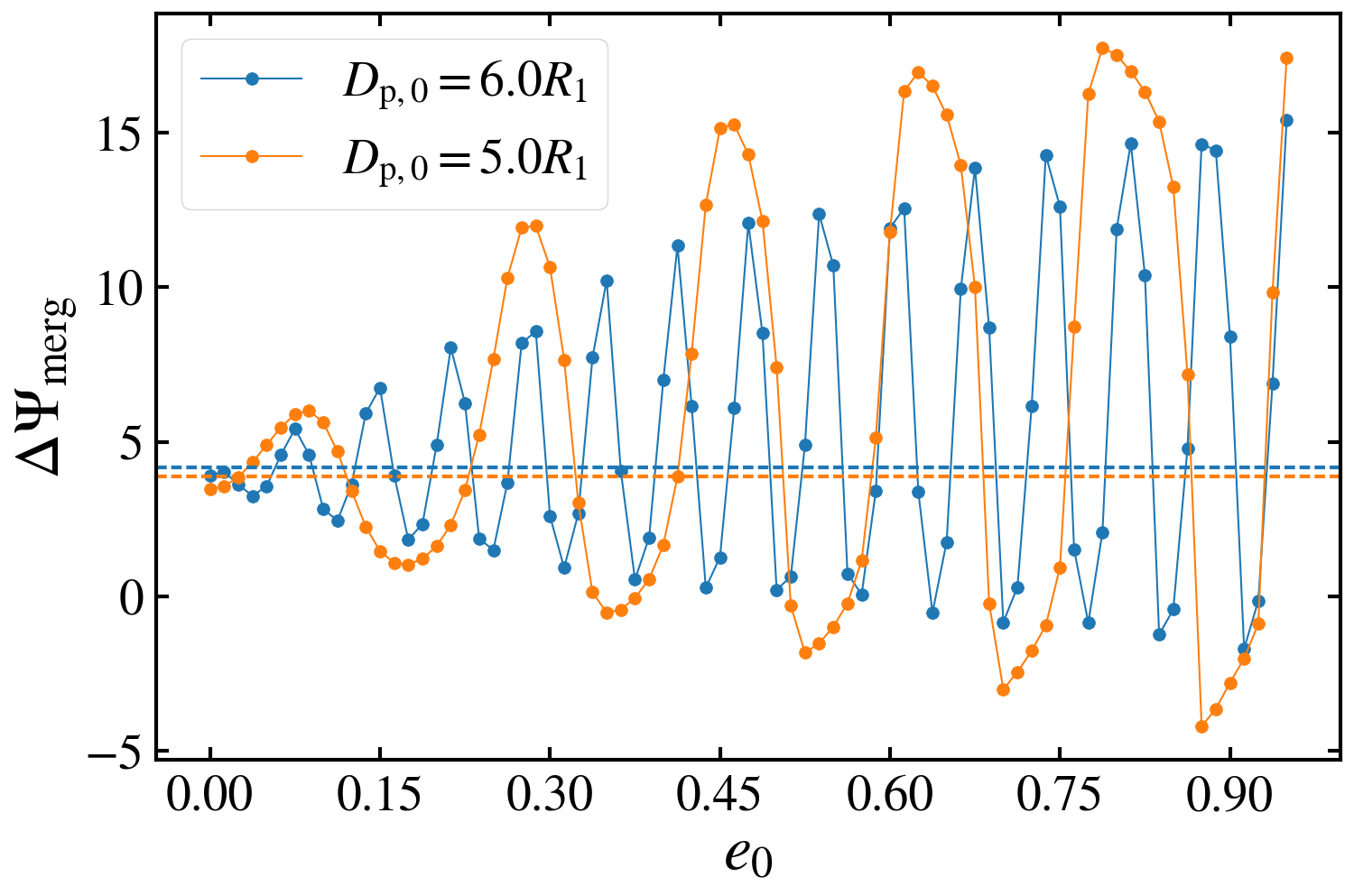}
	\caption{Cumulative GW phase difference between a calculation with dynamical tides and without at $t_{\rm merg}$ (the time when $D = 2.5 R_1$) as a function of initial eccentricity $e_0$ for two different values of the initial pericenter distance (see Fig.~\ref{fig:waveform}). The two dashed lines show the result for circular orbits (see Eq.~\ref{eq:DelPhiCirc}).}
	\label{fig:Psi_merg}
\end{figure*}

\begin{figure*}
	\includegraphics[width = 6in]{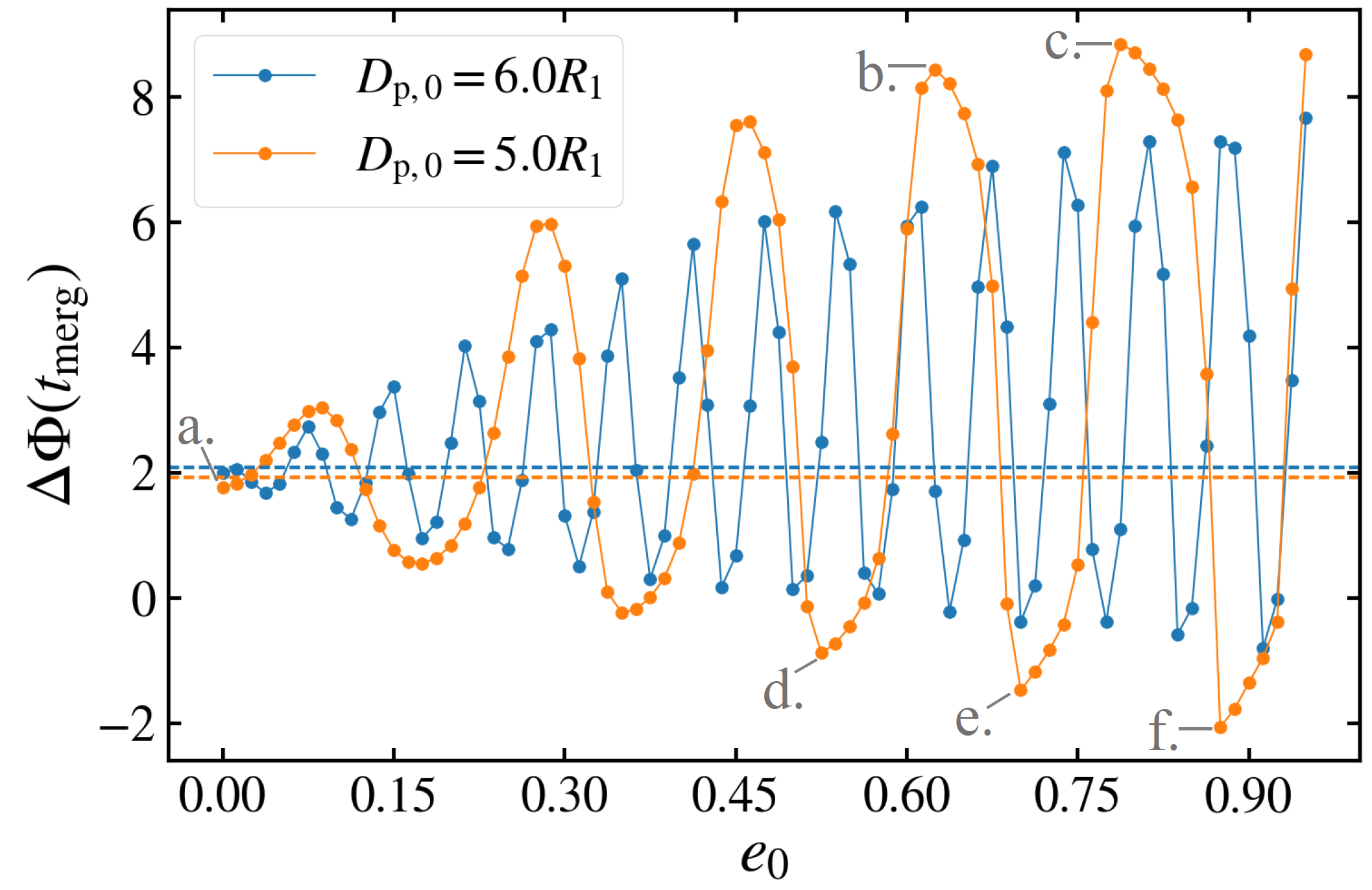}
	\includegraphics[width=6.5in]{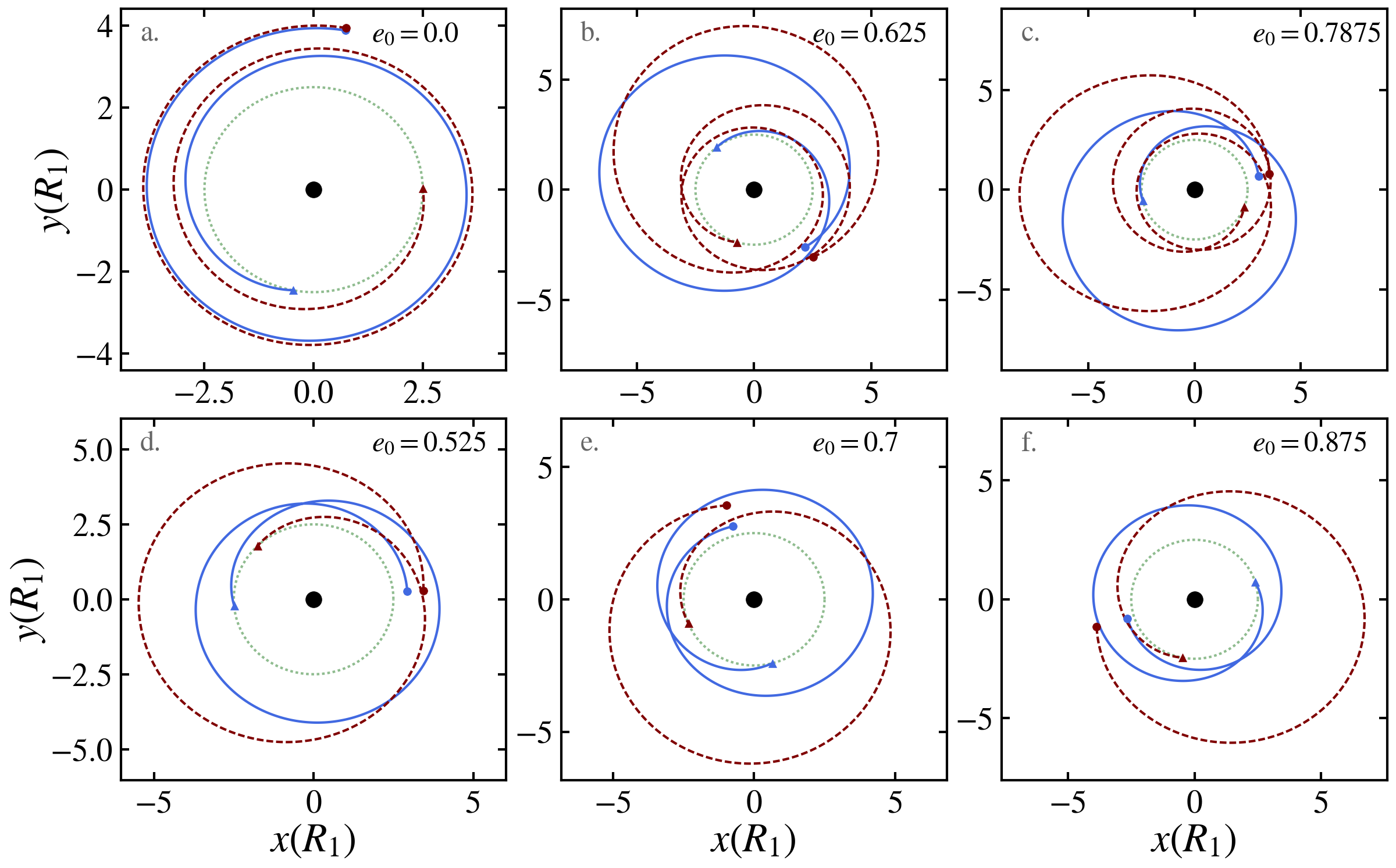}
	\caption{The upper panel is the same as in Fig.~\ref{fig:Psi_merg} with the orbital phase rather than the gravitational phase. The lower panels show the final few orbits for calculations with (blue solid lines) and without (red dashed lines) dynamical tides for binaries with $D_{\rm p,0} = 5.0 R_1$ and values of $e_0$ that maximize (b. and c.) or minimize (d., e., and f.) $\Delta \Phi_{\rm merg}$. The dotted green circles indicate $D = 2.5 R_1$.}
	\label{fig:Phi_merg}
\end{figure*}

We have integrated Eqs.~(\ref{eq:cdot}), (\ref{eq:Dddot}), and (\ref{eq:Phiddot}) for coalescing NS binaries on initially eccentric orbits and compared the results with integrations that do not include tidal effects [$c_\alpha(t) = 0$]. Our goal is to quantify how tides affect the orbital evolution of a coalescing NS binary and the resulting gravitational waveform. 

A sample integration is shown in Fig.~\ref{fig:OrbitalEvolution} for an equal mass $M_1=M_2=1.4M_\odot$, $R_1=R_2=10$~km NS binary with initial pericenter distance $D_{\rm p,0} = 6 R_1$ and initial eccentricity $e_0=0.9$. At time $t=0$, the NSs are at apocenter with separation $D_0 = D_{\rm p,0}(1+e_0)/(1-e_0)$, and $E_\alpha = 0$ for both NSs, i.e, there is no energy in the f-mode. We define $e_0$ in terms of the ratio of the initial angular velocity, $\dot{\Phi}_0$, to $\dot{\Phi}_{\rm circ, 0}$, the angular velocity required to maintain a circular orbit with radius $D_0$ (in the absence of tidal effects and gravitational radiation) such that 
\begin{equation}
\sqrt{1-e_0}\equiv \frac{\dot{\Phi}_0}{\dot{\Phi}_{\rm circ, 0}}. \label{eq:defe0}
\end{equation}
We obtain $\dot{\Phi}_{\rm circ, 0}$ by solving Eq.~(\ref{eq:Dddot}) for $\dot{\Phi}(t=0)$ using $D(0)=D_0$ and $\dot{D}(0) = \ddot{D}(0) = c_\alpha (0) = 0$.
The GW frequency at the initial pericenter passage is 
\begin{equation}
f_{\rm p,0} = \frac{1}{\pi}\sqrt{\frac{M_t (1+e_0)}{a_0(1-e_0)^3}}. \label{eq:fp0}
\end{equation}
For our sample system, $f_{\rm p,0} = 575$~Hz. We stop the integration when the binary separation $D(t)$ becomes smaller than $2.5 R_1$. The time when a system reaches this criterion is labeled $t_{\rm merg}$. 

Comparison of the calculations with and without tides reveals that tides typically speed up the binary coalescence (see Fig.~\ref{fig:OrbitalEvolution}). The difference in phase traversed before merger (related to the number of orbits completed between $t=0$ and $t=t_{\rm merg}$) is
\begin{equation}
\Delta \Phi(t) = \Phi_{\rm Ntide}(t) - \Phi_{\rm tide} (t), \label{eq:DeltaPhi}
\end{equation}
where $\Phi_{\rm tide}(t)$ [$\Phi_{\rm Ntide}$] is the orbital phase at time $t$ for a calculation that includes [does not include] tidal effects. For the example depicted in Fig.~\ref{fig:OrbitalEvolution}, we see that $\Delta \Phi$ reaches $4$ radians, mostly accumulated in the last $\sim 10$~ms prior to merger. The mode energy (Eq.~\ref{eq:defEmodes}) approaches $\sim 10^{-3}$ of the NS binding energy $(M_1^2/R_1)$ and is dominated by the $m=2$ prograde mode. 

To understand how tidal effects influence the GW signal, we calculate the waveform assuming that the binary is ``face-on." The components of the strain, $h_+$ and $h_\times$, are given by
\begin{align}
h_+ = \frac{1}{d}(\ddot{I}_{xx} - \ddot{I}_{yy}), && h_\times = \frac{2}{d}\ddot{I}_{xy},
\end{align}
where $I_{xx}$, $I_{yy}$, and $I_{xy}$ are components of the quadrupole moment tensor (the $xy$ coordinates are defined in the orbital plane), and $d$ is the distance to the system. Neglecting the quadrupole moment contributions from the oscillation modes, we have 
\begin{align}
I_{xx} = \mu D^2 \cos^2{\Phi}, && I_{yy} = \mu D^2 \sin^2{\Phi}, && I_{xy} = 2 \mu D^2 \sin{2\Phi}.
\end{align}
The waveform is given by
\begin{align}
h_+  = \frac{2 \mu}{d}&\left( \dot{D}^2 \cos{2 \Phi} + D \ddot{D} \cos{2 \Phi} - 4 D \dot{D} \dot{\Phi} \sin{2 \Phi} \right. \nonumber\\
&\left. -2D^2\dot{\Phi}^2 \cos{2 \Phi}-D^2\ddot{\Phi} \sin{2 \Phi}\right), \label{eq:hplus}\\
h_\times  = \frac{2\mu}{d}&\left(\dot{D}^2 \sin{2\Phi} + D \ddot{D} \sin {2\Phi} +4 D\dot{D}\dot{\Phi} \cos{2\Phi} \right. \nonumber\\
&\left.- 2 D^2 \dot{\Phi}^2\sin{2 \Phi} + D^2 \ddot{\Phi} \cos{ 2 \Phi}\right). \label{eq:hcross}
\end{align}
We can combine $h_+$ and $h_\times$ to form a complex strain with amplitude $A$ and phase $\Psi$,
\begin{equation}
A \text{e}^{-\text{i} \Psi(t)} = h_+ - i h_\times. \label{eq:APhidef}
\end{equation}
Figure~\ref{fig:waveform} shows the waveform that corresponds to the orbital evolution depicted in Fig.~\ref{fig:OrbitalEvolution}. The bottom panel shows the phase difference due to dynamical tides:
\begin{equation}
\Delta \Psi(t) = \Psi_{\rm Ntide}(t) - \Psi_{\rm tide}(t). \label{eq:DeltaPsi} 
\end{equation}
Note that $\Delta \Psi(t) \sim 2 \Delta \Phi(t)$, as one would expect from the form of Eqs.~(\ref{eq:hplus}) and (\ref{eq:hcross}). The spikes in $\Delta \Psi(t)$ occur as the system passes through pericenter. In the final stages of orbital decay and circularization, $\Delta \Psi$ quickly climbs. We label the value of $\Delta \Psi$ when $D= 2.5 R_1$ as $\Delta \Psi_{\rm merg}$ (see the bottom panel of Fig.~\ref{fig:waveform}).

Because the orbit is initially very eccentric, the orbital frequency sweeps through many (of order 10's of) resonances with the f-mode throughout orbital decay. The resonances occur when $\omega_\alpha = n \Omega_{\rm orb}$ for integer $n$ (see Section \ref{sec:noGR}). However, because of the rapid orbital decay and the large $n$ values involved, we do not see discrete resonant excitation of the mode amplitude (see Appendix). The orbital phase $\Phi$ does not suddenly increase when $\omega_\alpha = n \Omega_{\rm orb}$~\footnote{This behavior is different from the circular orbit case, where resonance with a low-frequency g-mode or r-mode occurs when $\omega_\alpha = 2 \Omega_{\rm orb}$ (see \cite{Xu18} and references therein).}.

We now quantify how the tidal effects on the waveform depend on $e_0$ and $D_{\rm p,0}$. Fig.~\ref{fig:Deltat} shows $\Delta t(N_p)$ (see the top panel of Fig.~\ref{fig:OrbitalEvolution}) as a function of the number of pericenter passages, $N_p$, for two different values of $D_{\rm p,0}$ and a handful of values for $e_0$. A large value of $e_0$ produces the largest timing shift for a given $D_{\rm p,0}$. A smaller value of $D_{\rm p,0}$ also yields a larger maximum value of $\Delta t$. For $D_{\rm p,0} = 5 R_1$, $\Delta t$ reaches 11~ms just before merger.  

We can also examine how the cumulative phase shift just before merger $\Delta \Psi_{\rm merg}$ (see Fig. \ref{fig:waveform}, lower panel) varies as a function of the initial eccentricity $e_0$. Fig.~\ref{fig:Psi_merg} shows that the excitation of the f-mode has the largest effect on systems that are highly eccentric and have small pericenter distances. 

Note that systems with larger $e_0$ do not fully circularize before $D=2.5 R_1$. As a result, $\Delta \Psi_{\rm merg}$ has a significant dependence on the orbital phase at merger. This effect is visible in the large oscillations in $\Delta \Psi_{\rm merg}$ as a function of $e_0$. Fig. \ref{fig:Phi_merg} depicts the orbit calculations for systems at the extrema of the $\Delta \Phi_{\rm merg}$ vs. $e_0$ curve to illustrate the reason for these oscillations. In general, dynamical tides remove energy from the binary orbit and enhance the rate of orbital decay. For a given $\Phi(t)$, the orbit-averaged separation (a similar concept to the semi-major axis) is always smaller for a calculation that includes dynamical tides than for one that does not. This is clearest in the circular case [see panel (a) of Fig.~\ref{fig:Phi_merg}], where the binary separation is always slightly smaller in the calculation that includes tides. For eccentric binaries, the binary separation at merger ($D=2.5 R_1$) can be significantly different from the orbit-averaged separation. Some binaries meet the condition for merger at pericenter and merge early at smaller $\Phi(t)$ and wider orbit-averaged separations than $2.5R_1$. Others meet the merger condition at apocenter and merge late at larger $\Phi(t)$. The maxima (minima) in the oscillations of $\Delta \Phi_{\rm merg}$ vs. $e_0$ occur when the calculation with tides results in merger at a relatively wide (tight) orbit while the calculation without tides leads to a merger at a tighter (wider) orbit. Panels (b) and (c) of Fig.~\ref{fig:Phi_merg} show calculations where $\Delta \Phi_{\rm merg}$ is maximized. Note that the final orbit with dynamical tides (blue solid line) is wider than the final orbit without dynamical tides (red dashed line) in these panels. Panels (d), (e), and (f) correspond to binaries where $\Delta \Phi_{\rm merg}$ is negative. For these systems, the final orbits are significantly wider for the calculations without dynamical tides than for those with tides included.

For small $e_0$, we can compare our $\Delta \Psi_{\rm merg}$ with the analytical result of the GW phase-shift due to tides. From Eq.~(66) of \cite{Lai94a}, the GW phase-shift due to the tidal distortion of $M_1$ (induced by $M_2$) as the binary decays from a semi-major axis of $a_i$ to $a_f$ is 
\begin{equation}
\Delta \Psi = \frac{3}{16}\kappa_n q_n\frac{R_1^5}{M_1^2M_t^{1/2}}\left(a_f^{-5/2}-a_i^{-5/2}\right)\left(\frac{39}{4} + \frac{M_t}{M_2}\right), \label{eq:DelPhiCircwDamping}
\end{equation}
where $q_n=(1-n/5)\kappa_n$, and $\kappa_n$ is defined in Eq.~(8) of \cite{Lai94a} \footnote{The usual tidal Love number is given by $k_2 = (3/2)\kappa_n q_n$.}. For a $\Gamma=2$ polytrope, $q_n = 4 \kappa_n/5$ and $\kappa_n=0.66$. The term proportional to $(M_t/M_2)$ in Eq.~(\ref{eq:DelPhiCircwDamping}) is due to the gravitational emission of the tidally forced f-mode. Our calculations do not account for this effect. Thus, for comparison with our numerical results, we use
\begin{equation}
\Delta \Psi' = \frac{117}{64}\kappa_n q_n\frac{R_1^5}{M_1^2M_t^{1/2}}\left(a_f^{-5/2}-a_i^{-5/2}\right), \label{eq:DelPhiCirc}
\end{equation}
which does not include GW emission associated with the mode. From Fig.~\ref{fig:Psi_merg}, our results for small $e_0$ are in agreement with the predicted value of $\Delta \Psi_{\rm merg}$ from Eq.~(\ref{eq:DelPhiCirc}). Note that, using Eq.~64 of \cite{Lai94a},
\begin{equation}
\Omega_{\rm orb} = \left(\frac{M_t}{a^3}\right)^{1/2}\left[1 + \frac{9}{4}\frac{\kappa_n q_n M_2}{M_1}\left(\frac{R_1}{a}\right)^5\right],
\end{equation}
Eq.~(\ref{eq:DelPhiCircwDamping}) is equivalent to
\begin{equation}
\frac{d \Psi}{d \ln f} = \frac{d \Psi_{\rm Ntide}}{d \ln f}
\left[1 - 3 \kappa_n q_n \left(\frac{R_1}{a}\right)^5\left(\frac{11 M_2}{M_1} + \frac{M_t}{M_1}\right)\right],\label{eq:DelPhivsFreq}
\end{equation}
where $d \Psi_{\rm Ntide}/d \ln f$ corresponds to the GW phase evolution without tidal effects. Equation (\ref{eq:DelPhivsFreq}) is the same as the expression derived in \cite{Flanagan08}.

\section{Discussion\label{sec:Discussion}}
We have demonstrated that dynamical tides (i.e. tidal excitations of NS f-modes) can have a significant
effect on the orbits and therefore the GW signals from eccentric CO binaries with at least one NS. We have developed a model that couples the evolution of the NS f-mode with 2.5PN orbital evolution to track changes in the ``f-mode+eccentric orbit" system as the orbit circularizes and moves toward coalescence. This model can be readily applied to a NS-NS binary or a BH-NS binary. In general, the transfer of energy from the orbit to the f-mode speeds up the orbital decay and advances the phase of the GW signal. We have used our model to quantify how the f-mode excitation affects the timing of peaks in the GW signal as the binary moves toward coalescence for systems with a range of initial pericenter distances and eccentricities. We have found that systems with large eccentricities (for a given pericenter) may experience GW phase shifts due to tides that are nearly an order of magnitude larger than the phase shift produced by a circular merger (see Fig.~\ref{fig:Psi_merg}).

Although the event rate of eccentric CO binary mergers with a NS is highly uncertain, such systems could offer a wealth of information on the equation of state of NSs. With some refinements, our model could be used to predict the timing of pericenter passages in eccentric NS binaries. To do this, our model would need to be modified to include the effects of NS spin, f-mode damping due to gravitational emission, and most importantly higher order PN effects. In particular, gravitational radiation associated with the tidally excited f-mode is coherent with the orbit, and can lead to a GW phase shift that is comparable to that from f-mode excitation (as computed in this paper). Therefore, our results provide a minimum expected phase shift due to dynamical tides in eccentric NS binaries. 

\appendix*
\section{Orbital decay through f-mode resonances}\label{Appendix}
\begin{figure}
	\includegraphics[width=\columnwidth]{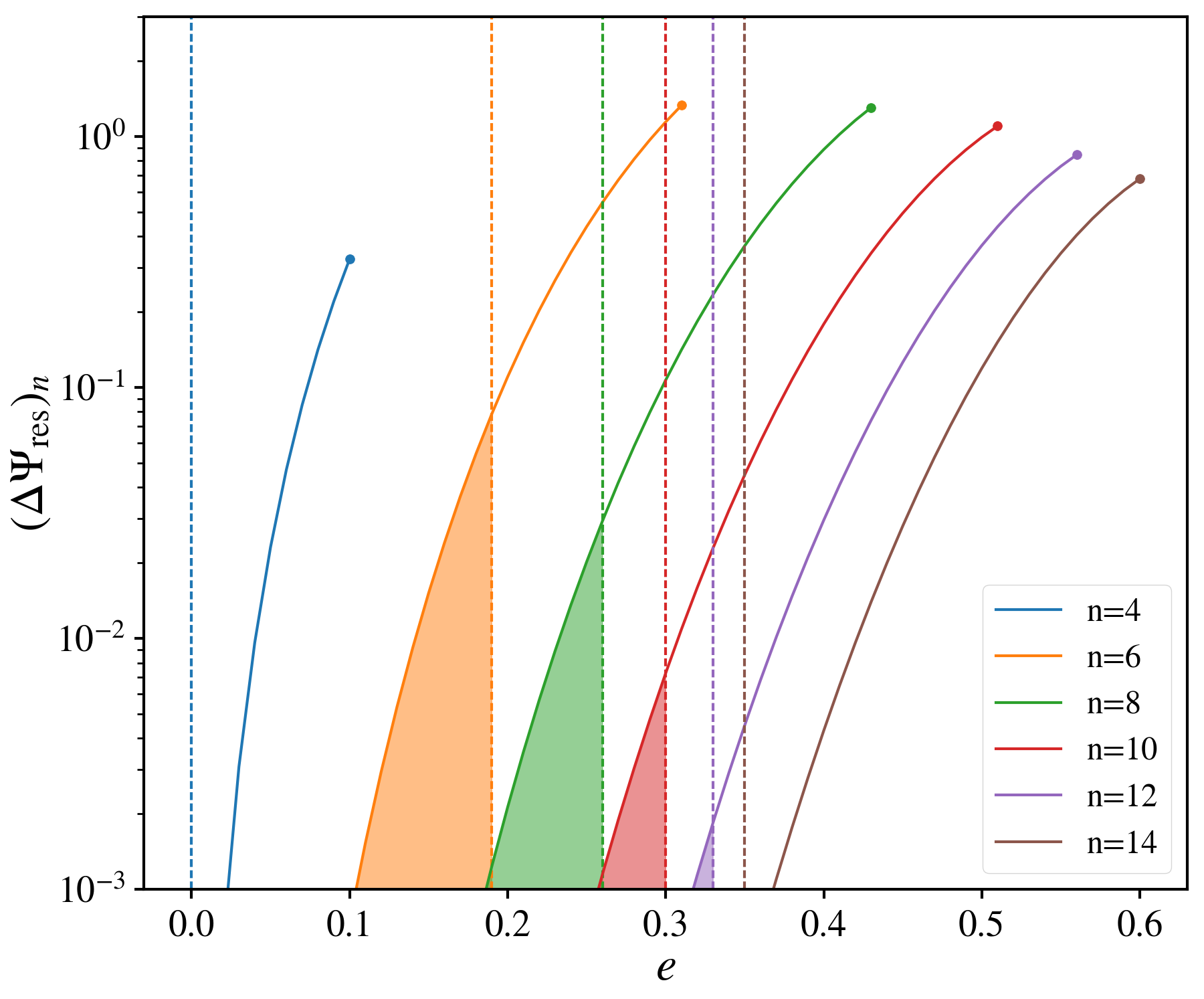}
	\caption{The GW phase shift $(\Delta \Psi_{\rm res})_n=2(\Delta \Phi_{\rm res})_n$ (see Eq.~\ref{eq:DeltaPsiRes}) due to mode-orbit resonance ($\omega_\alpha = n \Omega_{\rm orb}$ with positive integer $n$) as a function of the orbital eccentricity at the resonance. The results are for the $l=m=2$ f-mode of a $\Gamma=2$ polytropic NS model with $M_1 = 1.4~M_\odot$ and $R_1 = 10~\text{km}$ in an equal mass binary. The dashed lines indicate $\delta N_{\rm res}= 1$ (see Eq.~\ref{eq:deltaNres}). The results for $(\Delta \Psi_{\rm res})_n$ (solid lines) are valid when $\delta N_{\rm res}>1$, in the shaded region to the left of the dashed lines. The maximum displayed $e$ for each value of $n$ corresponds to the condition that the pericenter distance $D_{\rm p}$ exceeds $2.5R_1$.}
	\label{fig:PsiRes}
\end{figure}
As the eccentric binary orbit decays due to gravitational radiation, the orbital frequency $\Omega_{\rm orb}$ encounters resonances with the f-mode frequency $\omega_\alpha$ such that 
\begin{equation}
\omega_\alpha = n \Omega_{\rm orb} \label{eq:resonance},
\end{equation}
with integer $n$ (see also Section \ref{sec:noGR}). 
When the orbital frequency sweeps through a resonance slowly (over the course of multiple orbits), the NS mode energy can increase significantly, resulting in enhanced orbital decay and a phase shift in the gravitational waveform. This phase shift has been calculated for a variety of NS models and oscillation modes in the case of a circular orbit \cite[e.g.][]{Lai94b,Shibata94,Reisenegger94,Ho99,Lai06,Yu17a,Yu17b,Andersson18,Xu18}, and recently was considered for the low eccentricity case ($e \ll 1$) near the $\omega_\alpha = 3 \Omega_{\rm orb}$ resonance \cite{Yang19}. We generalize this calculation to higher-order resonances and arbitrary eccentricities, and show that these resonances generally produce a small GW phase shift.

When the orbital decay due to gravitational radiation occurs on a much longer timescale than an orbital period, we can approximate the gravitational potential produced by $M_2$ on $M_1$ (Eq.~\ref{eq:potential}) as a sum of multiple forcing frequencies, $n\Omega_{\rm orb}$, with positive integer $n$. We neglect PN effects (other than gravitational radiation) in this analysis. The time evolution of the mode amplitude $c_\alpha$ satisfies \cite[e.g.][]{Fuller12a,Vick17} 
\begin{equation}
\dot{c}_\alpha + \text{i} \omega_\alpha c_\alpha = \frac{i M_2 W_{lm}Q_\alpha}{2\omega_\alpha a^{l+1}}\sum_n{F_{mn}}\exp{\left[-\text{i}\int^t dt\;n\Omega_{\rm orb}(t)\right]}, \label{eq:cdotsum}
\end{equation}
where $a$ is the Newtonian semi-major axis, and
\begin{equation}
F_{mn} = \frac{1}{\pi} \int_0^\pi \frac{\cos[n (E - e\sin{E})-m\Phi(t)]}{(1-e\cos{E})^2}dE, 
\end{equation}
with $E$ the eccentric anomaly and 
\begin{equation}
\cos{\Phi(t)} = \frac{\cos{E} - e}{1-e\cos{E}}.
\end{equation}
Note that $G=M_1=R_1=1$ in Eq.~(\ref{eq:cdotsum}). Solving for $c_\alpha(t)$ yields 
\begin{equation}
c_\alpha \text{e}^{\text{i}\omega_\alpha t} = \sum_n F_{mn}\int dt \frac{\text{i} M_2 W_{lm}Q_\alpha}{2\omega_\alpha a^{l+1}}\text{e}^{\text{i}\left[\omega_\alpha t - \int^t dt\;n\Omega_{\rm orb}(t)\right]}.
\end{equation}
As $\Omega_{\rm orb}$ increases due to orbital decay, large changes in the mode amplitude can occur when the orbit sweeps through a resonance with the mode frequency. 

If the orbital decay is sufficiently slow, the mode amplitude after an encounter with the resonance $\omega_\alpha = n \Omega_{\rm orb}$ can be evaluated using the stationary phase approximation, giving
\begin{equation}
|c_{\alpha}| \simeq \frac{\text{i} M_2 W_{lm}Q_\alpha F_{mn}}{2\omega_\alpha a_n^{l+1}} \left(\frac{2 \pi}{n \dot{\Omega}_{\rm orb,n}}\right)^{1/2}, \label{eq:Deltacres}
\end{equation}
with $a_n$ and $\dot{\Omega}_{\rm orb,n}$ evaluated at the resonance. The change in the mode energy is $(\Delta E_{\rm res})_n = 2 \omega_\alpha^2 |c_\alpha|^2$. The associated change in the orbital phase due to the resonance is given by \cite{Lai94b}
\begin{equation}
(\Delta \Phi_{\rm res})_n \simeq -\left(\Omega_{\rm orb} t_D \frac{\Delta E_{\rm res}}{|E_{\rm orb}|}\right)_n, \label{eq:DelPhiRes}
\end{equation} 
where $E_{\rm orb} = -M_1M_2/2a$ is the orbital energy, and $t_D=|a/\dot{a}|$ is the orbital decay time due to gravitational radiation. Using (see Eq. 5.6 of \cite{Peters64})
\begin{equation}
\frac{\dot{\Omega}_{\rm orb}}{\Omega_{\rm orb}} = \frac{3}{2t_D} = \frac{96}{5} \frac{M_1^3q(1+q)}{a^4}\mathcal{F}(e) \label{eq:Omegadot},
\end{equation}
where $q = M_2/M_1$, and
\begin{equation}
\mathcal{F}(e) \equiv \frac{1}{(1-e^2)^{7/2}}\left(1 + \frac{73}{24}e^2 + \frac{37}{96}e^4\right), 
\end{equation}
we find
\begin{equation}
(\Delta \Phi_{\rm res})_n \simeq \frac{- 25 \pi}{3\times 2^9}\left(\frac{R_1}{M_1}\right)^5\frac{n}{\bar{\omega}_\alpha^{2}q(1+q)}\left[\frac{W_{2m}Q_\alpha F_{mn}}{\mathcal{F}(e)}\right]^2, \label{eq:DeltaPsiRes}
\end{equation}
where $\bar{\omega} = \omega/(M_1R_1^{-3})^{1/2}$, and $e$ is the eccentricity at resonance.
The phase shift in the gravitational waveform is $(\Delta \Psi_{\rm res})_n \simeq 2 (\Delta \Phi_{\rm res})_n$.

Equations~(\ref{eq:Deltacres}), (\ref{eq:DelPhiRes}) and (\ref{eq:DeltaPsiRes}) are valid only when the orbital decay is sufficiently slow. From Eq.~(\ref{eq:Deltacres}), we see that the change in the mode amplitude corresponds to the rate of change of the mode amplitude at resonance multiplied by the duration of the resonance,
\begin{equation}
\delta t_{\rm res} \equiv \left(\frac{2 \pi}{n \dot{\Omega}_{\rm orb,n}}\right)^{1/2}.
\end{equation} 
The number of orbital cycles during which resonance occurs is
\begin{equation}
\delta N_{\rm res} = \frac{\Omega_{\rm orb,n}}{2\pi}\delta t_{\rm res} = \left[\frac{5}{192 \pi}\left(\frac{R_1}{M_1}\right)^{5/2}\frac{(1+q)^{1/3}}{q}\frac{n^{2/3}}{\bar{\omega}^{5/3}_\alpha\mathcal{F}(e)}\right]^{1/2}. \label{eq:deltaNres}
\end{equation}
Resonance is significant only when $\delta N_{\rm res} \gtrsim 1$. When $\delta N_{\rm res} \lesssim 1$, the orbit moves through the resonance too quickly to strongly excite the oscillation.

We calculate $(\Delta \Psi_{\rm res})_n$ for the f-mode of a $\Gamma=2$ polytropic NS model with $M_1 = 1.4~M_\odot$ and $R_1 = 10~\text{km}$ over a large range of $n$ and $e$ and display the results in Fig.~\ref{fig:PsiRes}. For NS binaries that satisfy $\delta N_{\rm res} \gtrsim 1$, we find that $(\Delta \Psi_{\rm res})_n$ is always less than 0.1. We conclude that for eccentric NS binaries, f-mode resonances do not contribute significantly to the tidally generated phase shift. This finding is consistent with our numerical integrations in Section~\ref{sec:Results}, which did not exhibit sudden increases in the mode energy corresponding to mode-orbit resonances.

\begin{acknowledgments}
	We thank Larry Kidder, Prayesh Kumar and Saul Teukolsky for useful discussions. This work is supported in part by the NSF grant AST-1715246. MV is supported by a NASA Earth and Space Sciences Fellowship in Astrophysics.
\end{acknowledgments}

\bibliography{References}

\begin{thebibliography}{64}
\expandafter\ifx\csname natexlab\endcsname\relax\def\natexlab#1{#1}\fi
\expandafter\ifx\csname bibnamefont\endcsname\relax
  \def\bibnamefont#1{#1}\fi
\expandafter\ifx\csname bibfnamefont\endcsname\relax
  \def\bibfnamefont#1{#1}\fi
\expandafter\ifx\csname citenamefont\endcsname\relax
  \def\citenamefont#1{#1}\fi
\expandafter\ifx\csname url\endcsname\relax
  \def\url#1{\texttt{#1}}\fi
\expandafter\ifx\csname urlprefix\endcsname\relax\def\urlprefix{URL }\fi
\providecommand{\bibinfo}[2]{#2}
\providecommand{\eprint}[2][]{\url{#2}}

\bibitem[{\citenamefont{{The LIGO Scientific Collaboration}
  et~al.}(2018)\citenamefont{{The LIGO Scientific Collaboration}, {the Virgo
  Collaboration}, {Abbott}, {Abbott}, {Abbott}, {Abraham}, {Acernese},
  {Ackley}, {Adams}, {Adhikari} et~al.}}]{Ligo18}
\bibinfo{author}{\bibnamefont{{The LIGO Scientific Collaboration}}},
  \bibinfo{author}{\bibnamefont{{the Virgo Collaboration}}},
  \bibinfo{author}{\bibfnamefont{B.~P.} \bibnamefont{{Abbott}}},
  \bibinfo{author}{\bibfnamefont{R.}~\bibnamefont{{Abbott}}},
  \bibinfo{author}{\bibfnamefont{T.~D.} \bibnamefont{{Abbott}}},
  \bibinfo{author}{\bibfnamefont{S.}~\bibnamefont{{Abraham}}},
  \bibinfo{author}{\bibfnamefont{F.}~\bibnamefont{{Acernese}}},
  \bibinfo{author}{\bibfnamefont{K.}~\bibnamefont{{Ackley}}},
  \bibinfo{author}{\bibfnamefont{C.}~\bibnamefont{{Adams}}},
  \bibinfo{author}{\bibfnamefont{R.~X.} \bibnamefont{{Adhikari}}},
  \bibnamefont{et~al.}, \bibinfo{journal}{arXiv e-prints}
  (\bibinfo{year}{2018}), \eprint{1811.12907}.

\bibitem[{\citenamefont{{Abbott} et~al.}(2017)\citenamefont{{Abbott}, {Abbott},
  {Abbott}, {Acernese}, {Ackley}, {Adams}, {Adams}, {Addesso}, {Adhikari},
  {Adya} et~al.}}]{Ligo17}
\bibinfo{author}{\bibfnamefont{B.~P.} \bibnamefont{{Abbott}}},
  \bibinfo{author}{\bibfnamefont{R.}~\bibnamefont{{Abbott}}},
  \bibinfo{author}{\bibfnamefont{T.~D.} \bibnamefont{{Abbott}}},
  \bibinfo{author}{\bibfnamefont{F.}~\bibnamefont{{Acernese}}},
  \bibinfo{author}{\bibfnamefont{K.}~\bibnamefont{{Ackley}}},
  \bibinfo{author}{\bibfnamefont{C.}~\bibnamefont{{Adams}}},
  \bibinfo{author}{\bibfnamefont{T.}~\bibnamefont{{Adams}}},
  \bibinfo{author}{\bibfnamefont{P.}~\bibnamefont{{Addesso}}},
  \bibinfo{author}{\bibfnamefont{R.~X.} \bibnamefont{{Adhikari}}},
  \bibinfo{author}{\bibfnamefont{V.~B.} \bibnamefont{{Adya}}},
  \bibnamefont{et~al.}, \bibinfo{journal}{Physical Review Letters}
  \textbf{\bibinfo{volume}{119}}, \bibinfo{eid}{161101} (\bibinfo{year}{2017}),
  \eprint{1710.05832}.

\bibitem[{\citenamefont{{Lipunov} et~al.}(1997)\citenamefont{{Lipunov},
  {Postnov}, and {Prokhorov}}}]{Lipunov97}
\bibinfo{author}{\bibfnamefont{V.~M.} \bibnamefont{{Lipunov}}},
  \bibinfo{author}{\bibfnamefont{K.~A.} \bibnamefont{{Postnov}}},
  \bibnamefont{and} \bibinfo{author}{\bibfnamefont{M.~E.}
  \bibnamefont{{Prokhorov}}}, \bibinfo{journal}{Astronomy Letters}
  \textbf{\bibinfo{volume}{23}}, \bibinfo{pages}{492} (\bibinfo{year}{1997}).

\bibitem[{\citenamefont{{Lipunov} et~al.}(2017)\citenamefont{{Lipunov},
  {Kornilov}, {Gorbovskoy}, {Buckley}, {Tiurina}, {Balanutsa}, {Kuznetsov},
  {Greiner}, {Vladimirov}, {Vlasenko} et~al.}}]{Lipunov17}
\bibinfo{author}{\bibfnamefont{V.~M.} \bibnamefont{{Lipunov}}},
  \bibinfo{author}{\bibfnamefont{V.}~\bibnamefont{{Kornilov}}},
  \bibinfo{author}{\bibfnamefont{E.}~\bibnamefont{{Gorbovskoy}}},
  \bibinfo{author}{\bibfnamefont{D.~A.~H.} \bibnamefont{{Buckley}}},
  \bibinfo{author}{\bibfnamefont{N.}~\bibnamefont{{Tiurina}}},
  \bibinfo{author}{\bibfnamefont{P.}~\bibnamefont{{Balanutsa}}},
  \bibinfo{author}{\bibfnamefont{A.}~\bibnamefont{{Kuznetsov}}},
  \bibinfo{author}{\bibfnamefont{J.}~\bibnamefont{{Greiner}}},
  \bibinfo{author}{\bibfnamefont{V.}~\bibnamefont{{Vladimirov}}},
  \bibinfo{author}{\bibfnamefont{D.}~\bibnamefont{{Vlasenko}}},
  \bibnamefont{et~al.}, \bibinfo{journal}{Monthly Notices of the RAS}
  \textbf{\bibinfo{volume}{465}}, \bibinfo{pages}{3656} (\bibinfo{year}{2017}),
  \eprint{1605.01607}.

\bibitem[{\citenamefont{{Podsiadlowski}
  et~al.}(2003)\citenamefont{{Podsiadlowski}, {Rappaport}, and
  {Han}}}]{Podsiadlowski03}
\bibinfo{author}{\bibfnamefont{P.}~\bibnamefont{{Podsiadlowski}}},
  \bibinfo{author}{\bibfnamefont{S.}~\bibnamefont{{Rappaport}}},
  \bibnamefont{and} \bibinfo{author}{\bibfnamefont{Z.}~\bibnamefont{{Han}}},
  \bibinfo{journal}{Monthly Notices of the RAS} \textbf{\bibinfo{volume}{341}},
  \bibinfo{pages}{385} (\bibinfo{year}{2003}), \eprint{astro-ph/0207153}.

\bibitem[{\citenamefont{{Belczynski} et~al.}(2010)\citenamefont{{Belczynski},
  {Dominik}, {Bulik}, {O'Shaughnessy}, {Fryer}, and {Holz}}}]{Belczynski10}
\bibinfo{author}{\bibfnamefont{K.}~\bibnamefont{{Belczynski}}},
  \bibinfo{author}{\bibfnamefont{M.}~\bibnamefont{{Dominik}}},
  \bibinfo{author}{\bibfnamefont{T.}~\bibnamefont{{Bulik}}},
  \bibinfo{author}{\bibfnamefont{R.}~\bibnamefont{{O'Shaughnessy}}},
  \bibinfo{author}{\bibfnamefont{C.}~\bibnamefont{{Fryer}}}, \bibnamefont{and}
  \bibinfo{author}{\bibfnamefont{D.~E.} \bibnamefont{{Holz}}},
  \bibinfo{journal}{Astrophys. J., Lett.} \textbf{\bibinfo{volume}{715}},
  \bibinfo{pages}{L138} (\bibinfo{year}{2010}), \eprint{1004.0386}.

\bibitem[{\citenamefont{{Dominik} et~al.}(2012)\citenamefont{{Dominik},
  {Belczynski}, {Fryer}, {Holz}, {Berti}, {Bulik}, {Mandel}, and
  {O'Shaughnessy}}}]{Dominik12}
\bibinfo{author}{\bibfnamefont{M.}~\bibnamefont{{Dominik}}},
  \bibinfo{author}{\bibfnamefont{K.}~\bibnamefont{{Belczynski}}},
  \bibinfo{author}{\bibfnamefont{C.}~\bibnamefont{{Fryer}}},
  \bibinfo{author}{\bibfnamefont{D.~E.} \bibnamefont{{Holz}}},
  \bibinfo{author}{\bibfnamefont{E.}~\bibnamefont{{Berti}}},
  \bibinfo{author}{\bibfnamefont{T.}~\bibnamefont{{Bulik}}},
  \bibinfo{author}{\bibfnamefont{I.}~\bibnamefont{{Mandel}}}, \bibnamefont{and}
  \bibinfo{author}{\bibfnamefont{R.}~\bibnamefont{{O'Shaughnessy}}},
  \bibinfo{journal}{\apj} \textbf{\bibinfo{volume}{759}}, \bibinfo{eid}{52}
  (\bibinfo{year}{2012}), \eprint{1202.4901}.

\bibitem[{\citenamefont{{Dominik} et~al.}(2013)\citenamefont{{Dominik},
  {Belczynski}, {Fryer}, {Holz}, {Berti}, {Bulik}, {Mandel}, and
  {O'Shaughnessy}}}]{Dominik13}
\bibinfo{author}{\bibfnamefont{M.}~\bibnamefont{{Dominik}}},
  \bibinfo{author}{\bibfnamefont{K.}~\bibnamefont{{Belczynski}}},
  \bibinfo{author}{\bibfnamefont{C.}~\bibnamefont{{Fryer}}},
  \bibinfo{author}{\bibfnamefont{D.~E.} \bibnamefont{{Holz}}},
  \bibinfo{author}{\bibfnamefont{E.}~\bibnamefont{{Berti}}},
  \bibinfo{author}{\bibfnamefont{T.}~\bibnamefont{{Bulik}}},
  \bibinfo{author}{\bibfnamefont{I.}~\bibnamefont{{Mandel}}}, \bibnamefont{and}
  \bibinfo{author}{\bibfnamefont{R.}~\bibnamefont{{O'Shaughnessy}}},
  \bibinfo{journal}{\apj} \textbf{\bibinfo{volume}{779}}, \bibinfo{eid}{72}
  (\bibinfo{year}{2013}), \eprint{1308.1546}.

\bibitem[{\citenamefont{{Dominik} et~al.}(2015)\citenamefont{{Dominik},
  {Berti}, {O'Shaughnessy}, {Mandel}, {Belczynski}, {Fryer}, {Holz}, {Bulik},
  and {Pannarale}}}]{Dominik15}
\bibinfo{author}{\bibfnamefont{M.}~\bibnamefont{{Dominik}}},
  \bibinfo{author}{\bibfnamefont{E.}~\bibnamefont{{Berti}}},
  \bibinfo{author}{\bibfnamefont{R.}~\bibnamefont{{O'Shaughnessy}}},
  \bibinfo{author}{\bibfnamefont{I.}~\bibnamefont{{Mandel}}},
  \bibinfo{author}{\bibfnamefont{K.}~\bibnamefont{{Belczynski}}},
  \bibinfo{author}{\bibfnamefont{C.}~\bibnamefont{{Fryer}}},
  \bibinfo{author}{\bibfnamefont{D.~E.} \bibnamefont{{Holz}}},
  \bibinfo{author}{\bibfnamefont{T.}~\bibnamefont{{Bulik}}}, \bibnamefont{and}
  \bibinfo{author}{\bibfnamefont{F.}~\bibnamefont{{Pannarale}}},
  \bibinfo{journal}{\apj} \textbf{\bibinfo{volume}{806}}, \bibinfo{eid}{263}
  (\bibinfo{year}{2015}), \eprint{1405.7016}.

\bibitem[{\citenamefont{{Belczynski} et~al.}(2016)\citenamefont{{Belczynski},
  {Holz}, {Bulik}, and {O'Shaughnessy}}}]{Belczynski16}
\bibinfo{author}{\bibfnamefont{K.}~\bibnamefont{{Belczynski}}},
  \bibinfo{author}{\bibfnamefont{D.~E.} \bibnamefont{{Holz}}},
  \bibinfo{author}{\bibfnamefont{T.}~\bibnamefont{{Bulik}}}, \bibnamefont{and}
  \bibinfo{author}{\bibfnamefont{R.}~\bibnamefont{{O'Shaughnessy}}},
  \bibinfo{journal}{\nat} \textbf{\bibinfo{volume}{534}}, \bibinfo{pages}{512}
  (\bibinfo{year}{2016}), \eprint{1602.04531}.

\bibitem[{\citenamefont{{Portegies Zwart} and {McMillan}}(2000)}]{PZ00}
\bibinfo{author}{\bibfnamefont{S.~F.} \bibnamefont{{Portegies Zwart}}}
  \bibnamefont{and} \bibinfo{author}{\bibfnamefont{S.~L.~W.}
  \bibnamefont{{McMillan}}}, \bibinfo{journal}{Astrophys. J., Lett.}
  \textbf{\bibinfo{volume}{528}}, \bibinfo{pages}{L17} (\bibinfo{year}{2000}),
  \eprint{astro-ph/9910061}.

\bibitem[{\citenamefont{{Miller} and {Hamilton}}(2002)}]{Miller02}
\bibinfo{author}{\bibfnamefont{M.~C.} \bibnamefont{{Miller}}} \bibnamefont{and}
  \bibinfo{author}{\bibfnamefont{D.~P.} \bibnamefont{{Hamilton}}},
  \bibinfo{journal}{\apj} \textbf{\bibinfo{volume}{576}}, \bibinfo{pages}{894}
  (\bibinfo{year}{2002}), \eprint{astro-ph/0202298}.

\bibitem[{\citenamefont{{Wen}}(2003)}]{Wen03}
\bibinfo{author}{\bibfnamefont{L.}~\bibnamefont{{Wen}}},
  \bibinfo{journal}{\apj} \textbf{\bibinfo{volume}{598}}, \bibinfo{pages}{419}
  (\bibinfo{year}{2003}), \eprint{astro-ph/0211492}.

\bibitem[{\citenamefont{{O'Leary} et~al.}(2009)\citenamefont{{O'Leary},
  {Kocsis}, and {Loeb}}}]{OLeary09}
\bibinfo{author}{\bibfnamefont{R.~M.} \bibnamefont{{O'Leary}}},
  \bibinfo{author}{\bibfnamefont{B.}~\bibnamefont{{Kocsis}}}, \bibnamefont{and}
  \bibinfo{author}{\bibfnamefont{A.}~\bibnamefont{{Loeb}}},
  \bibinfo{journal}{Monthly Notices of the RAS} \textbf{\bibinfo{volume}{395}},
  \bibinfo{pages}{2127} (\bibinfo{year}{2009}), \eprint{0807.2638}.

\bibitem[{\citenamefont{{Miller} and {Lauburg}}(2009)}]{Miller09}
\bibinfo{author}{\bibfnamefont{M.~C.} \bibnamefont{{Miller}}} \bibnamefont{and}
  \bibinfo{author}{\bibfnamefont{V.~M.} \bibnamefont{{Lauburg}}},
  \bibinfo{journal}{\apj} \textbf{\bibinfo{volume}{692}}, \bibinfo{pages}{917}
  (\bibinfo{year}{2009}), \eprint{0804.2783}.

\bibitem[{\citenamefont{{Antonini} and {Perets}}(2012)}]{Antonini12}
\bibinfo{author}{\bibfnamefont{F.}~\bibnamefont{{Antonini}}} \bibnamefont{and}
  \bibinfo{author}{\bibfnamefont{H.~B.} \bibnamefont{{Perets}}},
  \bibinfo{journal}{\apj} \textbf{\bibinfo{volume}{757}}, \bibinfo{eid}{27}
  (\bibinfo{year}{2012}), \eprint{1203.2938}.

\bibitem[{\citenamefont{{Rodriguez} et~al.}(2015)\citenamefont{{Rodriguez},
  {Morscher}, {Pattabiraman}, {Chatterjee}, {Haster}, and
  {Rasio}}}]{Rodriguez15}
\bibinfo{author}{\bibfnamefont{C.~L.} \bibnamefont{{Rodriguez}}},
  \bibinfo{author}{\bibfnamefont{M.}~\bibnamefont{{Morscher}}},
  \bibinfo{author}{\bibfnamefont{B.}~\bibnamefont{{Pattabiraman}}},
  \bibinfo{author}{\bibfnamefont{S.}~\bibnamefont{{Chatterjee}}},
  \bibinfo{author}{\bibfnamefont{C.-J.} \bibnamefont{{Haster}}},
  \bibnamefont{and} \bibinfo{author}{\bibfnamefont{F.~A.}
  \bibnamefont{{Rasio}}}, \bibinfo{journal}{Physical Review Letters}
  \textbf{\bibinfo{volume}{115}}, \bibinfo{eid}{051101} (\bibinfo{year}{2015}),
  \eprint{1505.00792}.

\bibitem[{\citenamefont{{Samsing}}(2018)}]{Samsing18}
\bibinfo{author}{\bibfnamefont{J.}~\bibnamefont{{Samsing}}},
  \bibinfo{journal}{\prd} \textbf{\bibinfo{volume}{97}}, \bibinfo{eid}{103014}
  (\bibinfo{year}{2018}), \eprint{1711.07452}.

\bibitem[{\citenamefont{{Silsbee} and {Tremaine}}(2017)}]{Silsbee17}
\bibinfo{author}{\bibfnamefont{K.}~\bibnamefont{{Silsbee}}} \bibnamefont{and}
  \bibinfo{author}{\bibfnamefont{S.}~\bibnamefont{{Tremaine}}},
  \bibinfo{journal}{\apj} \textbf{\bibinfo{volume}{836}}, \bibinfo{eid}{39}
  (\bibinfo{year}{2017}), \eprint{1608.07642}.

\bibitem[{\citenamefont{{Antonini} et~al.}(2017)\citenamefont{{Antonini},
  {Toonen}, and {Hamers}}}]{Antonini17}
\bibinfo{author}{\bibfnamefont{F.}~\bibnamefont{{Antonini}}},
  \bibinfo{author}{\bibfnamefont{S.}~\bibnamefont{{Toonen}}}, \bibnamefont{and}
  \bibinfo{author}{\bibfnamefont{A.~S.} \bibnamefont{{Hamers}}},
  \bibinfo{journal}{\apj} \textbf{\bibinfo{volume}{841}}, \bibinfo{eid}{77}
  (\bibinfo{year}{2017}), \eprint{1703.06614}.

\bibitem[{\citenamefont{{Liu} and {Lai}}(2018)}]{Liu18}
\bibinfo{author}{\bibfnamefont{B.}~\bibnamefont{{Liu}}} \bibnamefont{and}
  \bibinfo{author}{\bibfnamefont{D.}~\bibnamefont{{Lai}}},
  \bibinfo{journal}{\apj} \textbf{\bibinfo{volume}{863}}, \bibinfo{eid}{68}
  (\bibinfo{year}{2018}), \eprint{1805.03202}.

\bibitem[{\citenamefont{{Liu} and {Lai}}(2019)}]{Liu19}
\bibinfo{author}{\bibfnamefont{B.}~\bibnamefont{{Liu}}} \bibnamefont{and}
  \bibinfo{author}{\bibfnamefont{D.}~\bibnamefont{{Lai}}},
  \bibinfo{journal}{Monthly Notices of the RAS} \textbf{\bibinfo{volume}{483}},
  \bibinfo{pages}{4060} (\bibinfo{year}{2019}), \eprint{1809.07767}.

\bibitem[{\citenamefont{{Liu} et~al.}(2019)\citenamefont{{Liu}, {Lai}, and
  {Wang}}}]{Liu19b}
\bibinfo{author}{\bibfnamefont{B.}~\bibnamefont{{Liu}}},
  \bibinfo{author}{\bibfnamefont{D.}~\bibnamefont{{Lai}}}, \bibnamefont{and}
  \bibinfo{author}{\bibfnamefont{Y.-H.} \bibnamefont{{Wang}}},
  \bibinfo{journal}{arXiv e-prints} \bibinfo{eid}{arXiv:1905.00427}
  (\bibinfo{year}{2019}), \eprint{1905.00427}.

\bibitem[{\citenamefont{{Kochanek}}(1992)}]{Kochanek92}
\bibinfo{author}{\bibfnamefont{C.~S.} \bibnamefont{{Kochanek}}},
  \bibinfo{journal}{\apj} \textbf{\bibinfo{volume}{385}}, \bibinfo{pages}{604}
  (\bibinfo{year}{1992}).

\bibitem[{\citenamefont{{Bildsten} and {Cutler}}(1992)}]{Bildsten92}
\bibinfo{author}{\bibfnamefont{L.}~\bibnamefont{{Bildsten}}} \bibnamefont{and}
  \bibinfo{author}{\bibfnamefont{C.}~\bibnamefont{{Cutler}}},
  \bibinfo{journal}{\apj} \textbf{\bibinfo{volume}{400}}, \bibinfo{pages}{175}
  (\bibinfo{year}{1992}).

\bibitem[{\citenamefont{{Lai} et~al.}(1994{\natexlab{a}})\citenamefont{{Lai},
  {Rasio}, and {Shapiro}}}]{Lai94a}
\bibinfo{author}{\bibfnamefont{D.}~\bibnamefont{{Lai}}},
  \bibinfo{author}{\bibfnamefont{F.~A.} \bibnamefont{{Rasio}}},
  \bibnamefont{and} \bibinfo{author}{\bibfnamefont{S.~L.}
  \bibnamefont{{Shapiro}}}, \bibinfo{journal}{\apj}
  \textbf{\bibinfo{volume}{420}}, \bibinfo{pages}{811}
  (\bibinfo{year}{1994}{\natexlab{a}}), \eprint{astro-ph/9304027}.

\bibitem[{\citenamefont{{Lai} and {Wiseman}}(1996)}]{Lai96b}
\bibinfo{author}{\bibfnamefont{D.}~\bibnamefont{{Lai}}} \bibnamefont{and}
  \bibinfo{author}{\bibfnamefont{A.~G.} \bibnamefont{{Wiseman}}},
  \bibinfo{journal}{\prd} \textbf{\bibinfo{volume}{54}}, \bibinfo{pages}{3958}
  (\bibinfo{year}{1996}), \eprint{gr-qc/9609014}.

\bibitem[{\citenamefont{{Baumgarte} et~al.}(1998)\citenamefont{{Baumgarte},
  {Cook}, {Scheel}, {Shapiro}, and {Teukolsky}}}]{Baumgarte98}
\bibinfo{author}{\bibfnamefont{T.~W.} \bibnamefont{{Baumgarte}}},
  \bibinfo{author}{\bibfnamefont{G.~B.} \bibnamefont{{Cook}}},
  \bibinfo{author}{\bibfnamefont{M.~A.} \bibnamefont{{Scheel}}},
  \bibinfo{author}{\bibfnamefont{S.~L.} \bibnamefont{{Shapiro}}},
  \bibnamefont{and} \bibinfo{author}{\bibfnamefont{S.~A.}
  \bibnamefont{{Teukolsky}}}, \bibinfo{journal}{\prd}
  \textbf{\bibinfo{volume}{57}}, \bibinfo{pages}{7299} (\bibinfo{year}{1998}),
  \eprint{gr-qc/9709026}.

\bibitem[{\citenamefont{{Binnington} and {Poisson}}(2009)}]{Binnington09}
\bibinfo{author}{\bibfnamefont{T.}~\bibnamefont{{Binnington}}}
  \bibnamefont{and}
  \bibinfo{author}{\bibfnamefont{E.}~\bibnamefont{{Poisson}}},
  \bibinfo{journal}{\prd} \textbf{\bibinfo{volume}{80}}, \bibinfo{eid}{084018}
  (\bibinfo{year}{2009}), \eprint{0906.1366}.

\bibitem[{\citenamefont{{Damour} and {Nagar}}(2009)}]{Damour09}
\bibinfo{author}{\bibfnamefont{T.}~\bibnamefont{{Damour}}} \bibnamefont{and}
  \bibinfo{author}{\bibfnamefont{A.}~\bibnamefont{{Nagar}}},
  \bibinfo{journal}{\prd} \textbf{\bibinfo{volume}{80}}, \bibinfo{eid}{084035}
  (\bibinfo{year}{2009}), \eprint{0906.0096}.

\bibitem[{\citenamefont{{Ury{\={u}}} et~al.}(2009)\citenamefont{{Ury{\={u}}},
  {Limousin}, {Friedman}, {Gourgoulhon}, and {Shibata}}}]{Uryu09}
\bibinfo{author}{\bibfnamefont{K.}~\bibnamefont{{Ury{\={u}}}}},
  \bibinfo{author}{\bibfnamefont{F.}~\bibnamefont{{Limousin}}},
  \bibinfo{author}{\bibfnamefont{J.~L.} \bibnamefont{{Friedman}}},
  \bibinfo{author}{\bibfnamefont{E.}~\bibnamefont{{Gourgoulhon}}},
  \bibnamefont{and}
  \bibinfo{author}{\bibfnamefont{M.}~\bibnamefont{{Shibata}}},
  \bibinfo{journal}{\prd} \textbf{\bibinfo{volume}{80}}, \bibinfo{eid}{124004}
  (\bibinfo{year}{2009}), \eprint{0908.0579}.

\bibitem[{\citenamefont{{Penner} et~al.}(2011)\citenamefont{{Penner},
  {Andersson}, {Samuelsson}, {Hawke}, and {Jones}}}]{Penner11}
\bibinfo{author}{\bibfnamefont{A.~J.} \bibnamefont{{Penner}}},
  \bibinfo{author}{\bibfnamefont{N.}~\bibnamefont{{Andersson}}},
  \bibinfo{author}{\bibfnamefont{L.}~\bibnamefont{{Samuelsson}}},
  \bibinfo{author}{\bibfnamefont{I.}~\bibnamefont{{Hawke}}}, \bibnamefont{and}
  \bibinfo{author}{\bibfnamefont{D.~I.} \bibnamefont{{Jones}}},
  \bibinfo{journal}{\prd} \textbf{\bibinfo{volume}{84}}, \bibinfo{eid}{103006}
  (\bibinfo{year}{2011}), \eprint{1107.0669}.

\bibitem[{\citenamefont{{Ferrari} et~al.}(2012)\citenamefont{{Ferrari},
  {Gualtieri}, and {Maselli}}}]{Ferrari12}
\bibinfo{author}{\bibfnamefont{V.}~\bibnamefont{{Ferrari}}},
  \bibinfo{author}{\bibfnamefont{L.}~\bibnamefont{{Gualtieri}}},
  \bibnamefont{and}
  \bibinfo{author}{\bibfnamefont{A.}~\bibnamefont{{Maselli}}},
  \bibinfo{journal}{\prd} \textbf{\bibinfo{volume}{85}}, \bibinfo{eid}{044045}
  (\bibinfo{year}{2012}), \eprint{1111.6607}.

\bibitem[{\citenamefont{{Xu} and {Lai}}(2017)}]{Xu18}
\bibinfo{author}{\bibfnamefont{W.}~\bibnamefont{{Xu}}} \bibnamefont{and}
  \bibinfo{author}{\bibfnamefont{D.}~\bibnamefont{{Lai}}},
  \bibinfo{journal}{\prd} \textbf{\bibinfo{volume}{96}}, \bibinfo{eid}{083005}
  (\bibinfo{year}{2017}), \eprint{1708.01839}.

\bibitem[{\citenamefont{{Lai} et~al.}(1994{\natexlab{b}})\citenamefont{{Lai},
  {Rasio}, and {Shapiro}}}]{Lai94}
\bibinfo{author}{\bibfnamefont{D.}~\bibnamefont{{Lai}}},
  \bibinfo{author}{\bibfnamefont{F.~A.} \bibnamefont{{Rasio}}},
  \bibnamefont{and} \bibinfo{author}{\bibfnamefont{S.~L.}
  \bibnamefont{{Shapiro}}}, \bibinfo{journal}{\apj}
  \textbf{\bibinfo{volume}{437}}, \bibinfo{pages}{742}
  (\bibinfo{year}{1994}{\natexlab{b}}), \eprint{astro-ph/9404031}.

\bibitem[{\citenamefont{{Flanagan} and {Hinderer}}(2008)}]{Flanagan08}
\bibinfo{author}{\bibfnamefont{{\'E}.~{\'E}.} \bibnamefont{{Flanagan}}}
  \bibnamefont{and}
  \bibinfo{author}{\bibfnamefont{T.}~\bibnamefont{{Hinderer}}},
  \bibinfo{journal}{\prd} \textbf{\bibinfo{volume}{77}}, \bibinfo{eid}{021502}
  (\bibinfo{year}{2008}), \eprint{0709.1915}.

\bibitem[{\citenamefont{{Lai}}(1994)}]{Lai94b}
\bibinfo{author}{\bibfnamefont{D.}~\bibnamefont{{Lai}}},
  \bibinfo{journal}{Monthly Notices of the RAS} \textbf{\bibinfo{volume}{270}},
  \bibinfo{pages}{611} (\bibinfo{year}{1994}), \eprint{astro-ph/9404062}.

\bibitem[{\citenamefont{{Shibata}}(1994)}]{Shibata94}
\bibinfo{author}{\bibfnamefont{M.}~\bibnamefont{{Shibata}}},
  \bibinfo{journal}{\prd} \textbf{\bibinfo{volume}{50}}, \bibinfo{pages}{6297}
  (\bibinfo{year}{1994}).

\bibitem[{\citenamefont{{Reisenegger} and {Goldreich}}(1994)}]{Reisenegger94}
\bibinfo{author}{\bibfnamefont{A.}~\bibnamefont{{Reisenegger}}}
  \bibnamefont{and}
  \bibinfo{author}{\bibfnamefont{P.}~\bibnamefont{{Goldreich}}},
  \bibinfo{journal}{\apj} \textbf{\bibinfo{volume}{426}}, \bibinfo{pages}{688}
  (\bibinfo{year}{1994}).

\bibitem[{\citenamefont{{Ho} and {Lai}}(1999)}]{Ho99}
\bibinfo{author}{\bibfnamefont{W.~C.~G.} \bibnamefont{{Ho}}} \bibnamefont{and}
  \bibinfo{author}{\bibfnamefont{D.}~\bibnamefont{{Lai}}},
  \bibinfo{journal}{Monthly Notices of the RAS} \textbf{\bibinfo{volume}{308}},
  \bibinfo{pages}{153} (\bibinfo{year}{1999}), \eprint{astro-ph/9812116}.

\bibitem[{\citenamefont{{Lai} and {Wu}}(2006)}]{Lai06}
\bibinfo{author}{\bibfnamefont{D.}~\bibnamefont{{Lai}}} \bibnamefont{and}
  \bibinfo{author}{\bibfnamefont{Y.}~\bibnamefont{{Wu}}},
  \bibinfo{journal}{\prd} \textbf{\bibinfo{volume}{74}}, \bibinfo{eid}{024007}
  (\bibinfo{year}{2006}), \eprint{astro-ph/0604163}.

\bibitem[{\citenamefont{{Yu} and {Weinberg}}(2017{\natexlab{a}})}]{Yu17a}
\bibinfo{author}{\bibfnamefont{H.}~\bibnamefont{{Yu}}} \bibnamefont{and}
  \bibinfo{author}{\bibfnamefont{N.~N.} \bibnamefont{{Weinberg}}},
  \bibinfo{journal}{Monthly Notices of the RAS} \textbf{\bibinfo{volume}{464}},
  \bibinfo{pages}{2622} (\bibinfo{year}{2017}{\natexlab{a}}),
  \eprint{1610.00745}.

\bibitem[{\citenamefont{{Yu} and {Weinberg}}(2017{\natexlab{b}})}]{Yu17b}
\bibinfo{author}{\bibfnamefont{H.}~\bibnamefont{{Yu}}} \bibnamefont{and}
  \bibinfo{author}{\bibfnamefont{N.~N.} \bibnamefont{{Weinberg}}},
  \bibinfo{journal}{Monthly Notices of the RAS} \textbf{\bibinfo{volume}{470}},
  \bibinfo{pages}{350} (\bibinfo{year}{2017}{\natexlab{b}}),
  \eprint{1705.04700}.

\bibitem[{\citenamefont{{Andersson} and {Ho}}(2018)}]{Andersson18}
\bibinfo{author}{\bibfnamefont{N.}~\bibnamefont{{Andersson}}} \bibnamefont{and}
  \bibinfo{author}{\bibfnamefont{W.~C.~G.} \bibnamefont{{Ho}}},
  \bibinfo{journal}{\prd} \textbf{\bibinfo{volume}{97}}, \bibinfo{eid}{023016}
  (\bibinfo{year}{2018}), \eprint{1710.05950}.

\bibitem[{\citenamefont{{Yang}}(2019)}]{Yang19}
\bibinfo{author}{\bibfnamefont{H.}~\bibnamefont{{Yang}}},
  \bibinfo{journal}{arXiv e-prints} \bibinfo{eid}{arXiv:1904.11089}
  (\bibinfo{year}{2019}), \eprint{1904.11089}.

\bibitem[{\citenamefont{{Chirenti} et~al.}(2017)\citenamefont{{Chirenti},
  {Gold}, and {Miller}}}]{Chirenti17}
\bibinfo{author}{\bibfnamefont{C.}~\bibnamefont{{Chirenti}}},
  \bibinfo{author}{\bibfnamefont{R.}~\bibnamefont{{Gold}}}, \bibnamefont{and}
  \bibinfo{author}{\bibfnamefont{M.~C.} \bibnamefont{{Miller}}},
  \bibinfo{journal}{\apj} \textbf{\bibinfo{volume}{837}}, \bibinfo{eid}{67}
  (\bibinfo{year}{2017}), \eprint{1612.07097}.

\bibitem[{\citenamefont{{Parisi} and {Sturani}}(2018)}]{Parisi18}
\bibinfo{author}{\bibfnamefont{A.}~\bibnamefont{{Parisi}}} \bibnamefont{and}
  \bibinfo{author}{\bibfnamefont{R.}~\bibnamefont{{Sturani}}},
  \bibinfo{journal}{\prd} \textbf{\bibinfo{volume}{97}}, \bibinfo{eid}{043015}
  (\bibinfo{year}{2018}), \eprint{1705.04751}.

\bibitem[{\citenamefont{{Yang} et~al.}(2018)\citenamefont{{Yang}, {East},
  {Paschalidis}, {Pretorius}, and {Mendes}}}]{Yang18}
\bibinfo{author}{\bibfnamefont{H.}~\bibnamefont{{Yang}}},
  \bibinfo{author}{\bibfnamefont{W.~E.} \bibnamefont{{East}}},
  \bibinfo{author}{\bibfnamefont{V.}~\bibnamefont{{Paschalidis}}},
  \bibinfo{author}{\bibfnamefont{F.}~\bibnamefont{{Pretorius}}},
  \bibnamefont{and} \bibinfo{author}{\bibfnamefont{R.~F.~P.}
  \bibnamefont{{Mendes}}}, \bibinfo{journal}{\prd}
  \textbf{\bibinfo{volume}{98}}, \bibinfo{eid}{044007} (\bibinfo{year}{2018}),
  \eprint{1806.00158}.

\bibitem[{\citenamefont{{Chaurasia} et~al.}(2018)\citenamefont{{Chaurasia},
  {Dietrich}, {Johnson-McDaniel}, {Ujevic}, {Tichy}, and
  {Br{\"u}gmann}}}]{Chaurasia18}
\bibinfo{author}{\bibfnamefont{S.~V.} \bibnamefont{{Chaurasia}}},
  \bibinfo{author}{\bibfnamefont{T.}~\bibnamefont{{Dietrich}}},
  \bibinfo{author}{\bibfnamefont{N.~K.} \bibnamefont{{Johnson-McDaniel}}},
  \bibinfo{author}{\bibfnamefont{M.}~\bibnamefont{{Ujevic}}},
  \bibinfo{author}{\bibfnamefont{W.}~\bibnamefont{{Tichy}}}, \bibnamefont{and}
  \bibinfo{author}{\bibfnamefont{B.}~\bibnamefont{{Br{\"u}gmann}}},
  \bibinfo{journal}{\prd} \textbf{\bibinfo{volume}{98}}, \bibinfo{eid}{104005}
  (\bibinfo{year}{2018}).

\bibitem[{\citenamefont{{Schenk} et~al.}(2002)\citenamefont{{Schenk}, {Arras},
  {Flanagan}, {Teukolsky}, and {Wasserman}}}]{Schenk02}
\bibinfo{author}{\bibfnamefont{A.~K.} \bibnamefont{{Schenk}}},
  \bibinfo{author}{\bibfnamefont{P.}~\bibnamefont{{Arras}}},
  \bibinfo{author}{\bibfnamefont{{\'E}.~{\'E}.} \bibnamefont{{Flanagan}}},
  \bibinfo{author}{\bibfnamefont{S.~A.} \bibnamefont{{Teukolsky}}},
  \bibnamefont{and}
  \bibinfo{author}{\bibfnamefont{I.}~\bibnamefont{{Wasserman}}},
  \bibinfo{journal}{\prd} \textbf{\bibinfo{volume}{65}}, \bibinfo{eid}{024001}
  (\bibinfo{year}{2002}), \eprint{gr-qc/0101092}.

\bibitem[{\citenamefont{{Fuller} and {Lai}}(2012)}]{Fuller12a}
\bibinfo{author}{\bibfnamefont{J.}~\bibnamefont{{Fuller}}} \bibnamefont{and}
  \bibinfo{author}{\bibfnamefont{D.}~\bibnamefont{{Lai}}},
  \bibinfo{journal}{Monthly Notices of the RAS} \textbf{\bibinfo{volume}{420}},
  \bibinfo{pages}{3126} (\bibinfo{year}{2012}), \eprint{1107.4594}.

\bibitem[{\citenamefont{{Blanchet}}(2014)}]{Blanchet14}
\bibinfo{author}{\bibfnamefont{L.}~\bibnamefont{{Blanchet}}},
  \bibinfo{journal}{Living Reviews in Relativity}
  \textbf{\bibinfo{volume}{17}}, \bibinfo{eid}{2} (\bibinfo{year}{2014}),
  \eprint{1310.1528}.

\bibitem[{\citenamefont{{Loutrel} et~al.}(2019)\citenamefont{{Loutrel},
  {Liebersbach}, {Yunes}, and {Cornish}}}]{Loutrel19}
\bibinfo{author}{\bibfnamefont{N.}~\bibnamefont{{Loutrel}}},
  \bibinfo{author}{\bibfnamefont{S.}~\bibnamefont{{Liebersbach}}},
  \bibinfo{author}{\bibfnamefont{N.}~\bibnamefont{{Yunes}}}, \bibnamefont{and}
  \bibinfo{author}{\bibfnamefont{N.}~\bibnamefont{{Cornish}}},
  \bibinfo{journal}{Classical and Quantum Gravity}
  \textbf{\bibinfo{volume}{36}}, \bibinfo{eid}{025004} (\bibinfo{year}{2019}),
  \eprint{1810.03521}.

\bibitem[{\citenamefont{{Lincoln} and {Will}}(1990)}]{Lincoln90}
\bibinfo{author}{\bibfnamefont{C.~W.} \bibnamefont{{Lincoln}}}
  \bibnamefont{and} \bibinfo{author}{\bibfnamefont{C.~M.}
  \bibnamefont{{Will}}}, \bibinfo{journal}{\prd} \textbf{\bibinfo{volume}{42}},
  \bibinfo{pages}{1123} (\bibinfo{year}{1990}).

\bibitem[{\citenamefont{{Kidder} et~al.}(1993)\citenamefont{{Kidder}, {Will},
  and {Wiseman}}}]{Kidder93}
\bibinfo{author}{\bibfnamefont{L.~E.} \bibnamefont{{Kidder}}},
  \bibinfo{author}{\bibfnamefont{C.~M.} \bibnamefont{{Will}}},
  \bibnamefont{and} \bibinfo{author}{\bibfnamefont{A.~G.}
  \bibnamefont{{Wiseman}}}, \bibinfo{journal}{\prd}
  \textbf{\bibinfo{volume}{47}}, \bibinfo{pages}{R4183} (\bibinfo{year}{1993}),
  \eprint{gr-qc/9211025}.

\bibitem[{\citenamefont{{Fuller} and {Lai}}(2011)}]{Fuller11}
\bibinfo{author}{\bibfnamefont{J.}~\bibnamefont{{Fuller}}} \bibnamefont{and}
  \bibinfo{author}{\bibfnamefont{D.}~\bibnamefont{{Lai}}},
  \bibinfo{journal}{Monthly Notices of the RAS} \textbf{\bibinfo{volume}{412}},
  \bibinfo{pages}{1331} (\bibinfo{year}{2011}), \eprint{1009.3316}.

\bibitem[{\citenamefont{{Mardling}}(1995)}]{Mardling95a}
\bibinfo{author}{\bibfnamefont{R.~A.} \bibnamefont{{Mardling}}},
  \bibinfo{journal}{\apj} \textbf{\bibinfo{volume}{450}}, \bibinfo{pages}{722}
  (\bibinfo{year}{1995}).

\bibitem[{\citenamefont{{Lai}}(1996)}]{Lai96a}
\bibinfo{author}{\bibfnamefont{D.}~\bibnamefont{{Lai}}},
  \bibinfo{journal}{Astrophys. J., Lett.} \textbf{\bibinfo{volume}{466}}, \bibinfo{pages}{L35}
  (\bibinfo{year}{1996}), \eprint{astro-ph/9605096}.

\bibitem[{\citenamefont{{Ivanov} and {Papaloizou}}(2004)}]{IP04}
\bibinfo{author}{\bibfnamefont{P.~B.} \bibnamefont{{Ivanov}}} \bibnamefont{and}
  \bibinfo{author}{\bibfnamefont{J.~C.~B.} \bibnamefont{{Papaloizou}}},
  \bibinfo{journal}{Monthly Notices of the RAS} \textbf{\bibinfo{volume}{347}},
  \bibinfo{pages}{437} (\bibinfo{year}{2004}), \eprint{astro-ph/0303669}.

\bibitem[{\citenamefont{{Vick} and {Lai}}(2018)}]{Vick18}
\bibinfo{author}{\bibfnamefont{M.}~\bibnamefont{{Vick}}} \bibnamefont{and}
  \bibinfo{author}{\bibfnamefont{D.}~\bibnamefont{{Lai}}},
  \bibinfo{journal}{Monthly Notices of the RAS} \textbf{\bibinfo{volume}{476}},
  \bibinfo{pages}{482} (\bibinfo{year}{2018}), \eprint{1708.09392}.

\bibitem[{\citenamefont{{Wu}}(2018)}]{Wu18}
\bibinfo{author}{\bibfnamefont{Y.}~\bibnamefont{{Wu}}}, \bibinfo{journal}{Astron. J.}
  \textbf{\bibinfo{volume}{155}}, \bibinfo{eid}{118} (\bibinfo{year}{2018}),
  \eprint{1710.02542}.

\bibitem[{\citenamefont{{Vick} et~al.}(2019)\citenamefont{{Vick}, {Lai}, and
  {Anderson}}}]{Vick19}
\bibinfo{author}{\bibfnamefont{M.}~\bibnamefont{{Vick}}},
  \bibinfo{author}{\bibfnamefont{D.}~\bibnamefont{{Lai}}}, \bibnamefont{and}
  \bibinfo{author}{\bibfnamefont{K.~R.} \bibnamefont{{Anderson}}},
  \bibinfo{journal}{Monthly Notices of the RAS} \textbf{\bibinfo{volume}{484}},
  \bibinfo{pages}{5645} (\bibinfo{year}{2019}), \eprint{1812.05618}.

\bibitem[{\citenamefont{{Vick} et~al.}(2017)\citenamefont{{Vick}, {Lai}, and
  {Fuller}}}]{Vick17}
\bibinfo{author}{\bibfnamefont{M.}~\bibnamefont{{Vick}}},
  \bibinfo{author}{\bibfnamefont{D.}~\bibnamefont{{Lai}}}, \bibnamefont{and}
  \bibinfo{author}{\bibfnamefont{J.}~\bibnamefont{{Fuller}}},
  \bibinfo{journal}{Monthly Notices of the RAS} \textbf{\bibinfo{volume}{468}},
  \bibinfo{pages}{2296} (\bibinfo{year}{2017}).

\bibitem[{\citenamefont{Peters}(1964)}]{Peters64}
\bibinfo{author}{\bibfnamefont{P.~C.} \bibnamefont{Peters}},
  \bibinfo{journal}{Phys. Rev.} \textbf{\bibinfo{volume}{136}}
  (\bibinfo{year}{1964}).

\end{thebibliography}

\end{document}